\documentclass{article}
\usepackage{amsmath}

\usepackage[figuresright]{rotating}

\usepackage{times}
\usepackage{bm}
\usepackage[numbers]{natbib}

\usepackage{graphics,color,pict2e}
\usepackage{epsfig}
\usepackage{xcolor}

\date{January 26, 2026}

\begin{document}

\title{Vaccine Efficacy Estimands Implied by Common Estimators Used in Individual Randomized Field Trials}

\author{Michael P. Fay$^1$, Dean Follmann, Bruce J. Swihart, and Lauren E. Dang$^2$ \\[4pt]
\textit{Biostatistics Research Branch, National Institute of Allergy and Infectious Diseases (MPF, DF, LED)}, \\
\textit{and NIH Clinical Center (BJS), Rockville, Maryland, USA }
\\[2pt]
{mfay@niaid.nih.gov}}

\markboth%
{Fay, Follmann, Swihart, and Dang}
{Vaccine Efficacy Estimands}

\maketitle

\addtocounter{footnote}{1}
\footnotetext{Correspondence to M.P. Fay.}
\addtocounter{footnote}{1}
\footnotetext{Work of L.E. Dang was primarily done while at NIAID.}

\begin{abstract}
We review vaccine efficacy (VE) estimands for
susceptibility
in individual randomized trials with natural (unmeasured) exposure, where individual responses are measured as time from vaccination until an event (e.g., disease from the infectious agent). Common VE estimands are written as $1-\theta$, where $\theta$ is some ratio effect measure (e.g., ratio of  incidence rates, cumulative incidences, hazards, or odds) comparing 
outcomes under vaccination versus control.
Although the ratio effects are approximately equal with low control event rates, we explore the quality of that approximation using a nonparametric formulation.
Traditionally, the primary endpoint VE estimands  are full immunization (or biological) estimands that represent a subset of the intent-to-treat population, excluding those that have the event before the vaccine has been able to ramp-up to its full effect, requiring  care for proper causal interpretation.
Besides these primary VE estimands that summarize an effect of the vaccine over the full course of the study, we also consider local VE estimands that measure the
effect at particular time points.
We discuss interpretational difficulties of local VE estimands (e.g., depletion of susceptibles bias), and using frailty models as sensitivity analyses for the individual-level causal effects over time.

\end{abstract}


\maketitle

\section{Introduction}

The term {\it vaccine efficacy } (VE) is used in the literature to describe a variety of protective effects of a vaccine on a population \citep{halloran2010design}.
For each protective effect of interest, we can define a causal VE estimand that is the answer to a specific question about the effect of the vaccine on a particular outcome by a specified timepoint, in a given target population.
Table 1 of \citet{halloran1997study}  suggest a two-way classification of these VE estimands.
In one direction, VE estimands are classified into either effects in an exposed population (conditional estimands) or effects among a target population regardless of exposure status (unconditional estimands).
In the other direction (i.e., for each of the conditional or unconditional types) there are 3 types of estimands: VE for susceptibility (protection of the vaccinee from disease or infection),
VE for infectiousness (reduction in the ability of the vaccinee to infect others), and VE
combining the other two.
Besides these VE estimands, others have been proposed that measure VE on transmission risk, such as the VE for prevalent infection
\citep{rinta2009estimation,lipsitch2021interpreting,follmann2023vaccine}.

These various VE estimands are useful for understanding different aspects of how a vaccine is working in
a population.
In this paper, we focus only on unconditional (to exposure) VE estimands for susceptibility.
This class of VE estimands are often
estimated from
a randomized trial of naturally exposed individuals  to measure the effect of the vaccine to protect the vaccinee from disease by the infectious agent of interest \citep[e.g., the VEs estimated from large COVID-19 vaccine trials such as:][]{thomas2021safety,baden2021efficacy,sadoff2021safety,heath2021safety}.
We further restrict our focus to VE estimands that may be represented as a functional of the distributions of the time to first event
under test or control vaccination.
All of these VE estimands may be written as $VE = 1 - \theta$, where $\theta$ is a ratio effect such as the: incidence rate ratio, cumulative incidence ratio, hazard ratio,  cumulative hazard ratio, or  odds ratio.
We further classify these estimands as cumulative ones (measuring the VE over the full course of the study) or local ones (measuring the VE at specific times since vaccination).

We consider first cumulative VE estimands, which are candidates for the primary endpoint of a randomized vaccine trial. Our focus is on choosing among the different VE estimands for this purpose.
Section~\ref{sec-ideal.characteristics} reviews several ideal characteristics desired of an estimand
from a randomized trial, and applies them to vaccine trials.
In Section~\ref{sec-defining} we introduce potential outcome notation
as we will formally address the causal inference desired from the trial.
In Section~\ref{sec-cumulative}  we individually introduce different cumulative VE estimands
from an intent-to-treat perspective.
In practice, often VE estimators are used without explicitly defining the estimand first.
So, we begin with common estimators and then present the associated causal estimand that each estimator consistently estimates
under an ideal randomized trial.
We define these estimands nonparametrically as a function of time
since randomization,
so that the estimands are not tied to any specific model. 
Although the estimands are not tied to any specific model, it is useful to define
a `constancy model' for each estimand, which is the model such that the true value of the estimand does not change when evaluated at different timepoints (e.g., one minus cumulative incidence ratio at 6 months versus at 1 year).
For example,
the proportional hazards model is the constancy model for the ratio estimand that is consistently
estimated by the hazard ratio from the binary
intervention
Cox model (see equation~\ref{eq:VE.Cox.weighted.mean.of.hazards}). 
Our atypical
nonparametric formulation allows us to prespecify an estimand without knowing if its constancy model holds,
 and
also  allows us to compare the different VE estimands under different models and different lengths of trials as we do in Section~\ref{sec-comparison.of.estimands.general.case}.

Although
the
intention-to-treat
perspective
is the default
for many treatment trials \citep{gupta2011intention}, it is rarely
the primary endpoint
perspective
for vaccine trials. For vaccine trials, the primary endpoint usually represents
a full immunization effect (also called a biological effect, \citep{horne2001intent}), representing the effect of the vaccine
if taken as directed and after it has had a ramp-up period to allow the vaccinees' immune system to become fully effective. Thus,
to approximate the desired causal effect, 
events are typically counted only after the ramp-up period, and individuals that have events prior to the end of the ramp-up period are removed from the risk set.
We review VE estimands associated with this strategy (assuming adherence to the vaccine schedule),
showing why there is no simple causal interpretation without strong assumptions,
since the conditional population that survived to the end of the ramp-up period may be different between arms and can lead to improper causal comparisons.
Others have addressed this ramp-up period issue \citep{gilbert2013sensitivity,michiels2022estimation}, but our approach using the nonparametric VE estimands is different (see Sections~\ref{sec-ramp-up}
and \ref{sec-parametric.ramp.up}).

Finally, we consider  the local VE estimands, where the primary interest is on how the vaccine changes the individual vaccinee's risk of having the event as time changes.
Unfortunately, in a randomized trial we only observe each individual's response under one intervention (either the control or the active vaccine). Thus, from a randomized trial the most easily measured effects are effects on the population-level, not the individual-level. Changing population-level effects over time as measured by the hazard ratio are difficult to interpret in terms of their effects on individuals \citep{hernan2010hazards,aalen2015causal,martinussen2020subtleties,fay2024causal}.
The interpretational problem is mainly due to
non-matching conditional populations,
since a hazard at time $t$ represents the instantaneous probability of an event occurring at $t$,  conditioning on surviving
to just before $t$. The difficulty is related to the heterogeneity of the population and the fact that the frailest most susceptible individuals tend to have the event before less frail individuals. If the vaccine works, then this heterogeneity can lead to a depletion of susceptibles bias where a vaccine that has a constant individual effect on the hazard ratio can have a declining population effect on the hazard ratio over time.
In Section~\ref{sec-heterogeneity} we review frailty models as a way to discuss heterogeneity of risk and the issues with interpretation of local VE estimands
on an individual level. In Section~\ref{sec-parametric.ramp.up} we combine a frailty model with
a piecewise Weibull model in order to address both local effects and heterogeneity.
Janvin and Stensrud \citep{janvin2025quantification} address vaccine waning by explicitly defining the causal estimand over time based on a controlled challenge, but
their estimand requires strong assumptions to be identified in a randomized field trial.

There are some other reviews on vaccine efficacy in the literature.
 \citet{michiels2022estimation}  have similarly focused on comparing different VE estimands that are primary endpoints in trials (see their Appendix B), but the focus of that paper was on comparing COVID-19 vaccines and all the differences in the primary endpoints (e.g., how the event was defined, differing dosing regimens), and our focus is  on general mathematical comparisons between the different risk measures that apply to vaccines for any disease. In fact, many of these issues may be applied to any randomized trial with a survival endpoint. Our focus is on
causal
 estimands, with much less focus on
estimators and the details related to them, such as algorithmic calculation methods, or methods for dealing with censoring.
We only discuss intercurrent events due to events in the ramp-up period, and do not explore the many other
types of intercurrent events (e.g., non-adherence to vaccine regimen) and ways of defining estimands to account for them
\citep[see][]{fu2023application,beckers2025adopting}.
For more wide-ranging reviews on all types of endpoints in vaccine trials, including Phase I and II trials and cluster randomized, see \citet{halloran1997study}, \citet{halloran2010design}, or \citet{hudgens2004endpoints}.



\section{Vaccine Efficacy Estimands: Some Ideal Characteristics}
\label{sec-ideal.characteristics}

Before we introduce precise mathematical notation, we discuss some ideal
characteristics of a
causal
 vaccine efficacy estimand that is the effect to be estimated by
the primary endpoint in an individually randomized controlled trial.
The estimand is usually a scalar,
but it can also be a
multidimensional or infinite dimensional parameter, such as a function of the marginal distributions that changes over time.
The estimand is the thing we are trying to estimate with our estimator (i.e., our analysis method).
The ICH E9(R1) guidance on estimands in clinical trials \citep{ICH.E9.R1.2021}  gives a definition of an estimand for a treatment trial, which we quote, except replacing ``vaccine'' for ``treatment'', and ``participants'' for ``patients'':
\begin{quote}
{\it A precise description of the [vaccine] effect reflecting the clinical question posed by the trial
objective. It summarizes at a population-level what the outcomes would be in the same [participants]
under different [vaccine] conditions being compared. }
\end{quote}
Although the quote does not mention `causal' inference explicitly, the second sentence implies a causal interpretation.
For this paper, we want to know whether the test vaccine is preventing or delaying the event (e.g., disease [or infection] of interest)
compared to the control vaccine in the population of interest.

We list some ideal characteristics desirable for any
causal
estimand from a randomized controlled trial, and comment on how those characteristics apply to a randomized vaccine trial.
\begin{description}
\item[Simplicity:] The VE estimand should be easy to explain to a lay audience.
\item[Robustness:] The VE estimand should be
well defined without strong model assumptions.
\item[Transportable:] The VE estimand should be reasonably applied to other situations and populations besides the specific study. This is especially important for aspects of the study that cannot be
    controlled by design, like the attack rate (e.g., the proportion of events in the control arm up to any specific time $t$), which for studies with event-based stopping rules affects the length of the study.
\item[Prespecified:] The VE estimand should be prespecified. Prespecification removes the possibility or the appearance that the VE estimand was chosen to show the test vaccine in the most positive light.
\item[Identifiable:] The VE estimand should be identifiable from the individual randomized trial (i.e., a consistent estimator of the VE estimand exists under reasonable assumptions).
\item[Consistent Estimator: ]
  The chosen estimator is consistent for the estimand under reasonable assumptions.
\item[Efficient Estimator:] Practically, we want there to be an
  estimator for the estimand that is reasonably efficient.
\item[Mechanistic:] The VE estimand is derived by a mechanistic model of how the vaccine works.
\end{description}

These 8 characteristics may compete with each other. For example, in order to have more robustness, the estimand may need to be less simple.
For VE estimands, one particularly difficult compromise is between transportability and robustness. For transportability, we want our VE estimand to not depend on the length of the study or the exposure rates, yet in order to achieve that,
 strong modeling assumptions must be true.
In this paper, we emphasize prespecification, and therefore define our estimands nonparametrically,
so that minimal model assumptions are needed for their definition.
For estimands used to tease out the mechanism of action of the vaccine, it may make sense to explore different models on the data and their associated estimands and use Occam's razor to pick the model that most simply explains the data. For example, \citet{halloran2010design} (Section 7.1.2)
reviewing \citet{smith1984assessment} suggest that if a model using cumulative incidence ratio does not change much over time, then this suggests an all-or-none vaccine mechanism, while if a model using
the incidence rate does not change much over time, then this suggests a leaky vaccine mechanism. This
is reasonable for hypothesizing about vaccine mechanism, but it is not always feasible for
choosing a robust primary endpoint for a randomized trial, where prespecification of the estimand is required.

Although it is somewhat backwards to talk about an estimator prior to the estimand, we do that with the hazard ratio from the Cox model because of its historical importance. Further, it supplies an example of the difficulty in weighing the different characteristics.
Consider a randomized trial that prespecifies the hazard ratio from the Cox model with only
an intervention
effect as its primary analysis.
 Details are given in Section~\ref{sec-cox}. Without censoring and without assuming proportional hazards, the Cox hazard ratio estimator is consistent for a complicated estimand
 that can be written as a ratio of weighted hazards, where the weights themselves depend on the estimand. That estimand violates the simplicity characteristic, unless
 we violate the robustness characteristic and assume the proportional hazards model, in which case it simplifies to a constant population hazard ratio estimand.
 That constant ratio may be easier to explain but will not accurately describe the estimand that we have estimated if the proportional hazards assumption is false.
  Under proportional hazards, the Cox hazard ratio estimator is efficient and consistent for the population hazard ratio estimand under independent censoring
  \citep[see e.g.,][]{andersen1993statistical}.
   But even the simpler population hazard ratio estimand does not have a straightforward causal interpretation \citep{fay2024causal}. Further, requiring
   a falsifiable assumption to achieve simpler interpretation of an estimand is risky. We see that
   choosing an estimand is difficult, and the user must decide how much to emphasize each of the 8 characteristics for each application.

\section{Potential Outcomes and Vaccine Efficacy Estimands}
\label{sec-defining}

Consider a two-arm randomized trial with $n$ participants
randomized to either the test vaccine arm
or the control vaccine arm. For the $i$th participant, let $Z_i$ be their intervention arm ($Z_i=0$ for control
and $Z_i=1$ for test), and let   $[T_i(0), T_i(1)]$ be their potential outcomes,
where $T_i(z)$ is the outcome they would have if randomized to arm $z=0,1$.
 For each participant we only observe one of their potential outcomes.  In this paper $T_i(z)$ is continuous (except in Appendix~\ref{sec-weighted.avg.discrete.haz}) and represents time from randomization until the first incidence of
the event (e.g., disease from the agent of interest). We assume throughout that
the first vaccination dose happens at randomization.
Because of staggered entry into the study, time $t$,
as measured from randomization,
 will not necessarily be at the same calendar time
for each individual.
Assume $T_i(z) \sim F_z$, where $F_z(t) = Pr \left[  T_i(z) \leq t \right]$, for $z=0,1$.
Throughout this manuscript, we focus on
 causal estimands that should be identifiable in an ideal study
and hence assume that we have no censoring and  unbiased outcome measurement. Because we discuss randomized trials of a well-defined intervention (active vaccination vs. control), the causal identification assumptions of exchangeability, positivity, and consistency (defined in \citet{hernan2024causal}) are satisfied.
We make the convenience assumption that there is  no interference between individuals, that the
potential outcomes for an individual do
not depend on the
intervention assignment
of other individuals.
For infectious diseases we know this is an unrealistic assumption; however, when the study population
is a small proportion of the population of potential infectious contacts, the non-interference assumption may be approximately true. Another reason not to worry about the no interference assumption is that the estimands we study here (direct unconditional ones, see next paragraph) often change little when exposure changes (see e.g., Section~\ref{sec-CH}), and the interference will predominately affect the exposure.
Since we are focusing on estimands not estimators, we will not address how to estimate $F_z$
(e.g., how to estimate it due to censoring caused by staggered accrual).

We consider some vaccine efficacy estimands listed as unconditional direct vaccine efficacy estimands given in Table~1 of \citet{halloran1997study}. They are ``unconditional'' because we do not condition on
exposure to infection, and they are ``direct'' meaning that the effect is due to the direct protection of the vaccinated participants, not due to the indirect effect of their neighbors being vaccinated.

Following Hern\'an and Robins \citep{hernan2024causal}, all of the VE estimands we study are population causal estimands because they can be expressed as functionals of the marginal distributions under the different study interventions
(i.e., as functionals of $F_0$ and $F_1$).
The issue with VE estimands is that they depend on the control incidence rates, which may be hard to predict precisely. Thus, we often use event-driven study designs, where the end of the study, $\tau$,
is not a fixed time but occurs after the total amount of events in both arms reaches a predetermined value. If $\tau$ is not fixed in advance, we assume that the event-driven stopping rule associated with $\tau$ is predetermined.

For cumulative VE estimands, we would ideally choose a VE estimand
for which the true value of the estimand
does not depend on the value of $\tau$ (for transportability) or depend on a specific
data generating model (for robustness); however, we cannot in general have both
being correctly specified.
Instead we define the cumulative VE estimands as a function of $\tau$ (representing the effect acting through all the time from $0$ to $\tau$), and for each of those VE estimands we give a constancy model: the model such that the VE estimand does not change with time.
For example, if the cumulative  VE estimand is $1-F_1(\tau)/F_0(\tau)$, then if we assume the constancy model where $F_1(t) \propto F_0(t)$
where $t$ is within some range that includes $\tau$,
then the true value of the VE estimand will not change if we change $\tau$ to another value within that range.
 Because we will define the VE estimands nonparametrically, we can choose the
 estimand
 where we suspect that
 its
 constancy model is approximately correct, but even if that model does not hold,
 the VE estimand will still be well defined.

\section{Intention-to-Treat Cumulative VE Estimands}
\label{sec-cumulative}

\subsection{Overview}

The ICH E9(R1) \citep{ICH.E9.R1.2021} recommends choosing the (causal) estimand first,
then deciding on the study
design, statistical model, its assumptions, and an estimator that allows consistent estimates of the causal estimand
 under identification assumptions.
 In defining the primary endpoint in traditional vaccine randomized trials, there has been less focus on the estimands and more on the estimators \citep[although that is changing, see e.g.,][]{fu2023application}.
Because of this, we begin our discussion with the estimators.
We list many common estimators and the
 causal
estimands to which they are consistent under certain models
and assumptions;
 however, we define the estimands based on $F_0$ and $F_1$ and the study length, $\tau$,
in order to study the estimands without needing a specific correctly specified model.
We review four traditional VE
 estimators and their implied
estimands for measuring cumulative effects from time 0 to $\tau$
and discuss one newer VE estimand based on  the cumulative hazard ratio estimand,
which is particularly useful for  robustly summarizing average hazard ratio effects.

For this section we study estimands that follow the intention-to-treat (ITT) principle. In other words, the ITT estimands relate to the population that includes all participants
randomized, and do not exclude from the population individuals based on post-randomization events \citep{gupta2011intention}.
For randomized vaccine trials the primary endpoint is often a full immunization effect estimand (see Section~\ref{sec-ramp-up}), but we start with  the simpler ITT VE estimands that have a more straightforward causal interpretation.

\subsection{Cumulative Incidence}
\label{sec-ITT.CI}

A common estimator of a ratio effect uses the ratio of cumulative incidences at the end of the study, which can be written in terms of the empirical distribution functions, $\hat{F}_1(\tau)$ and $\hat{F}_0(\tau)$:
\begin{eqnarray*}
\hat{\theta}_{CI}(\tau) & =  &
\frac{ \frac{\mbox{ number of events by $\tau$ in the vaccine arm}}{\mbox{number of people in the vaccine arm }} }{ \frac{\mbox{ number of events by $\tau$ in the control arm}}{\mbox{number of people in the control arm}} } \\
& = & \frac{\frac{1}{n_1} \sum_{i=1}^{n} I(T_i \leq \tau) I(Z_i=1) }{
\frac{1}{n_0} \sum_{i=1}^{n} I(T_i \leq \tau)I(Z_i=0)} = \frac{\hat{F}_1(\tau)}{\hat{F}_0(\tau)},
\end{eqnarray*}
where
$n_z$ is the number of people with $Z_i=z$,
 $T_i$ is the time to event for individual $i$, and $I(A)=1$ if $A$ is true and $0$ otherwise.  The value
$\hat{\theta}_{CI}(\tau)$ is a finite sample estimate of the ratio of cumulative incidences by time $\tau$ that we would observe if the whole target population of individuals (sometimes referred to as the super-population) were randomized to vaccine or placebo.

 By randomization and the assumptions mentioned in Section~\ref{sec-defining},
the statistical quantity we are estimating
 is equivalent to the causal estimand,
\[
\theta_{CI}(\tau)  =  \frac{Pr(T_i(1) \leq \tau )}{Pr(T_i(0) \leq \tau)} = \frac{F_1(\tau)}{F_0(\tau)},
\]
where $F_z(\tau)$ is sometimes called the {\it attack rate} for  arm $z$ \citep[see e.g.,][p. 21]{halloran2010design}.
 $\theta_{CI}(\tau)$ is 
the ratio of the proportion of the population who would have the outcome by time $\tau$ if everyone were vaccinated compared to the proportion of the population who would have the outcome by time $\tau$ if everyone were given placebo.
The cumulative incidence VE estimand is
$VE_{CI}(\tau) = 1-\theta_{CI}(\tau)$ and its estimator is $\widehat{VE}_{CI}(\tau)$.

Here is a simple explanation. Suppose
 ${VE}_{CI}(\tau)$ $=70\%$.
Then we could say that on average a vaccinated person would reduce their probability of getting the event by time $\tau$ by $70\%$. Or we could say that on average one is 
$\frac{1}{{\theta}_{CI}(\tau)}$ 
$=\frac{1}{0.30} = 3.33$ times more likely to have the event by time $\tau$ if unvaccinated than if vaccinated. However, a well-specified estimand should also incorporate information regarding the target population, dates of enrollment, and duration of follow-up. Thus, for a particular trial, we might say that on average a vaccinated person in the United States in 2020 would reduce their probability of getting the event by time $\tau$ by $70\%$.

For transportability, we assume,
often implicitly, 
 that the VE would be the same for different populations, time periods, control exposure rates, and study durations. This type of transportability requires a constancy model that allows any $F_0(t)$ and assumes that $F_1(t) = \theta_{CI} F_0(t)$ for all $t$.
A simple mechanistic motivation for the constancy model is as follows.
We assume that the vaccine is an all-or-none vaccine \citep[see e.g.,][Chapter 7]{halloran2010design}, in the following sense. Suppose there are only two classes of people: those that the vaccine works completely (i.e.,
once vaccinated they cannot get the disease of interest) and those that the vaccine fails completely (i.e., after vaccination they act the same as if they had never got the vaccine). Suppose that the vaccination works in $\psi$ proportion of the population. Letting $F_0(t)$ be the cumulative distribution in the control arm. Then the cumulative  distribution in the test arm is $F_1(t) =  (1-\psi) F_0(t)$, so that $\theta_{CI}(t) = 1-\psi$ and $VE_{CI}(t) = \psi$. For an all-or-none vaccine $VE_{CI}(\tau)$ does not depend on $\tau$,
regardless of the event distribution over time in the control arm.

Unfortunately, for many diseases the all-or-none mode of action is not very realistic. Often the amount of infectious agent exposure can overwhelm the vaccinee, so that
for some individuals they are protected if exposed at a small dose of the infectious agent, but are not protected at large doses; however, importantly, often vaccinees may have less disease severity under the large exposure dose than they would have had if they had gotten the control vaccination.
A different, perhaps more realistic,  interpretation of an all-or-none vaccine may approximately hold in some natural exposure situations. If
at each infectious contact the natural exposure doses are similar enough and if the infectious contacts
are spread out in time, then we may be able to approximately classify  vaccinated people into two groups, those that would be protected from disease from their first potentially infectious contact and those that would not be protected. This second interpretation holds
if the contacts were spaced out enough,
and if the cumulative dose over time from repeated contacts does not lead to more disease,
which may be reasonable because those that were protected from the first exposure may be
protected by all future exposures since each exposure increases their natural immunity.


\subsection{Incidence Rate}

Another common estimator of a ratio effect uses the ratio of incidence rates over the course of the study,
\begin{eqnarray*}
\hat{\theta}_{IR}(\tau) & =  & \frac{ \frac{\mbox{ number of events by $\tau$ in the vaccine arm}}{\mbox{person-years at risk by $\tau$ in the vaccine arm}} }{ \frac{\mbox{ number of events by $\tau$ in the control arm}}{\mbox{person-years by $\tau$ at risk in the control arm}} },
\end{eqnarray*}
where ``by
$\tau$'' denotes events that occur from time 0 up to and including time $\tau$.
In the case without censoring, people are at risk until they have their first event, so the number
of person-years at risk in arm $z$ is
\begin{eqnarray*}
PYR_z(\tau) & = & \sum_{i=1}^{n} I(Z_i=z)  \min \left\{ \tau, T_i \right\}.
\end{eqnarray*}
An estimator of the  incidence rate vaccine efficacy is
\begin{eqnarray*}
\hat{\theta}_{IR}(\tau) & = &  \frac{\frac{\sum_{i=1}^{n} I(T_i \leq \tau) I(Z_i=1)}{PYR_1(\tau)}}{\frac{\sum_{i=1}^{n} I(T_i \leq \tau) I(Z_i=0)}{PYR_0(\tau)}}.
\end{eqnarray*}
For
a causal
estimand, we write the expected value of the person-years at risk
for intervention $z$
 as
\begin{eqnarray*}
E \left\{ PYR_z(\tau) \right\} & = &  n_z
E \left\{ \min \left[ \tau, T_i(z) \right] \right\}
= n_z \mu_z(\tau)
\end{eqnarray*}
where $\mu_z(\tau)$ is the restricted mean survival time parameter at $\tau$
\citep[see][]{karrison1997use} and can be reexpressed as
\begin{eqnarray*}
\mu_z(\tau) & = & \int_0^{\tau} S_z(t) dt,
\end{eqnarray*}
where $S_z(t) =1-F_z(t)$ is the survival function for
intervention
 $z$.
By randomization, the expectation defining $\mu_z(\tau)$ is the same as if the
entire target population got intervention $z$.
Thus, the
causal
estimand associated with vaccine efficacy by incidence rates is
\begin{eqnarray}
VE_{IR}(\tau) & = & 1 - \theta_{IR}(\tau) = 1 - \frac{ \frac{F_1(\tau)}{\mu_1(\tau)}}{\frac{F_0(\tau)}{\mu_0(\tau)}}. \label{eq:VE.IR}
\end{eqnarray}

%
%

Suppose $VE_{IR}(\tau)=70\%$, then
a simple explanation
is that
on average, the incidence rate during the study is reduced by 70\% for
vaccinated compared to unvaccinated individuals, where incidence rate is the
expected number of events per person-years at risk.

A constancy model for $\theta_{IR}$ is the two-exponential model, where $F_0 \sim \textrm{Exponential}(\lambda_0)$ and
$F_1 \sim$ $\textrm{Exponential}(\lambda_1)$. Then $\theta_{IR}=\lambda_1/\lambda_0$, because
under the exponential distribution,  $S_z(t)= \exp( - \lambda_z t)$,
and the restricted mean survival time
for intervention
$z$ is
\begin{eqnarray*}
\mu_z(\tau) & = & \int_0^{\tau} S_z(t) dt = \int_0^{\tau} \exp( - \lambda_z t) dt \\
& = &  \frac{1}{\lambda_z} \left\{ 1 - \exp(-\lambda_z \tau) \right\} = \frac{ F_z(\tau) }{ \lambda_z },
\end{eqnarray*}
so that
\begin{eqnarray*}
\theta_{IR}(\tau) & = & \frac{ \frac{ F_1(\tau) }{ F_1(\tau)/\lambda_1 } }{ \frac{ F_0(\tau) }{ F_0(\tau)/\lambda_0 } } =  \frac{ \lambda_1 }{ \lambda_0}.
\end{eqnarray*}

Here is a mechanistic motivation for this model.
Assume the hazard rate (often called the force of infection, see \citet{halloran2010design} p. 26) for the control arm, $\lambda_0$, is in a steady state and not changing over time.
This means that both the probability of any at-risk individual getting the event in the next infinitesimal time period does not depend on the time $t$ since vaccination,
and that the nature of the population of people at risk at time $t$ does not change over time.
This assumption is violated when the exposure rate is changing with time, such as at the beginning of an epidemic.
Another violation of this assumption is if there is a frailty effect where certain individuals are more likely to have the event early, and hence be removed from the population at risk.
We discuss when the population is heterogeneous due to a frailty effect in Section~\ref{sec-heterogeneity}.
Next, we assume the vaccine is a leaky vaccine \citep[see e.g.,][p. 132-3]{halloran2010design}, meaning it acts the same on every individual by modifying the hazard of an event at any time by a constant, $\theta_{IR}$, then the
hazard
rate
in the vaccine arm is $\lambda_1 = \theta_{IR} \lambda_0$.  This results in an exponential distribution as well.

The two-exponential model is the simplest special parametric case of a proportional hazards model (which we define explicitly in the next section);
however, under the proportional hazards assumption when $\lambda_0(t)$ is not constant, then
$\theta_{IR}(t)$ is not necessarily constant.

\subsection{Cox Model}
\label{sec-cox}

Now consider the case where the ratio effect
estimator
is the hazard ratio estimator from the Cox model
\citep{cox1972regression,therneau2000modeling}. Specifically, consider the Cox model  where there is no censoring except at $\tau$, and, besides the baseline hazard, it has only one parameter which is associated with the
intervention
arm.
We write its hazard ratio estimator as
$\hat{\theta}_{Cox}(\tau)$, where the $\tau$ denotes that the estimator is calculated using all of the data from time $0$ up until the end of the study at time $\tau$. The Cox model can also be used to
estimate time varying hazard ratios, i.e., local hazard ratios at any time $t$,  $\theta_h(t)=\lambda_1(t)/\lambda_0(t)$, but
we will not discuss that until Section~\ref{sec-heterogeneity}.

In general (e.g., without assuming proportional hazards),
and under the identification assumptions of Section~\ref{sec-defining},
$\hat{\theta}_{Cox}(\tau)$ converges to the value of $\theta$ that solves \citep[see e.g.,][]{vansteelandt2024assumption}
\begin{eqnarray}
0 & = & \int_0^{\tau} \left\{ \frac{S_1(t) S_0(t) }{ \theta S_1(t) + S_0(t) } \right\} \left\{ \lambda_1(t) - \theta \lambda_0(t) \right\} dt,
\label{eq:CoxEstimand.from.leanCoxPaper}
\end{eqnarray}
where
$\lambda_z(t) = f_z(t)/S_z(t)$ and $f_z(t)=\partial F_z(t)/ \partial t$. 
We can reexpress equation~\ref{eq:CoxEstimand.from.leanCoxPaper} as solving for $\theta$
using a ratio of weighted mean hazard functions,
\begin{eqnarray}
\theta & = & \frac{ \int_0^{\tau} w(t; \theta) \lambda_1(t)  dt }{ \int_0^{\tau} w(t; \theta) \lambda_0(t) dt}
\label{eq:VE.Cox.weighted.mean.of.hazards}
\end{eqnarray}
where the weights themselves depend on $\theta$,
\begin{eqnarray}
w(t; \theta ) & = &  \frac{S_1(t) S_0(t) }{ \theta S_1(t) + S_0(t) }.
\label{eq:weights.for.Cox}
\end{eqnarray}
Let $\theta_{Cox}(\tau)$ be the $\theta$ that solves the above equations, then
\begin{eqnarray}
VE_{Cox}(\tau) & = & 1 - \theta_{Cox}(\tau).  \label{eq:VE.Cox}
\end{eqnarray}
We emphasize that the $\theta_{Cox}(t)$ is a  cumulative estimand giving a ratio of weighted hazards covering time from $0$ up to time $t$, not a local one like $\theta_h(t)$ that is only addressing the hazard ratio at time $t$.
Further,  $\theta_{Cox}(\tau)$ is defined
as in equation~(\ref{eq:CoxEstimand.from.leanCoxPaper})
when there is no censoring, but when there is independent censoring the expected value of the Cox model estimator of the hazard ratio depends on the censoring distribution (unless we have proportional hazards) \citep{struthers1986misspecified}.
We emphasize that the `Cox' subscript denotes the derivation based on the Cox model under a simplified ideal situation, but it is not the estimand that is consistently estimated by the Cox model in all situations.

It is hard to get a simple accurate description of $VE_{Cox}(\tau)$ when the
proportional hazard assumption does not hold
because any verbal description of the estimand must explain the complex weighting scheme that depends on the true value of the estimand itself,
although when events are rare $VE_{Cox}(\tau) \approx VE_{CI}(\tau) \approx VE_{IR}(\tau)$ (see Section~\ref{sec-comparison.of.estimands.general.case}).
When $F_0$ and $F_1$ are related by proportional hazards, then a  $VE_{Cox}(\tau)=70\%$
says a typical individual given the test vaccine will have a 70\% reduction in the
instantaneous event rate at any time post vaccination
compared to a typical individual given control vaccine.
It is generally not accurate to say that for any particular individual it reduces
their hazard rate by 70\%;
this is an individual-level interpretation which is not true without
strong untestable assumptions  (see Section~\ref{sec-heterogeneity} or \citep{fay2024causal}).
A simpler VE estimand that equals  $VE_{Cox}(\tau)$ under proportional hazards is
the one based on the cumulative hazard ratio (see Section~\ref{sec-CH}).

A semi-parametric constancy model for $\theta_{Cox}(t)$ is the proportional hazards model, where
$F_0$ and $F_1$ have proportional hazards such that for every $t$, $\lambda_1(t) = \theta_{Cox} \lambda_0(t)$,
and $\theta_{Cox}$ is constant.
The following are equivalent definitions of the proportional hazards assumption:
\begin{enumerate}
\item $\theta_{Cox} =\lambda_1(t)/\lambda_0(t)$ for every $t$, where $\lambda_z(\cdot)$ is the hazard function
    associated with  $F_z$ for $z=0,1$
\item $\theta_{Cox} = \log( S_1(t) )/ \log( S_0(t) )$ for every $t$,
\item $S_1(t) = \left\{ S_0(t) \right\}^{\theta_{Cox}}$ for every $t$,
\item
\label{PH:increasing.Transformation.to.Exponential}
 If $T_0 \sim F_0$ and $T_1 \sim F_1$ and there exists a strictly increasing transformation of the time scale, say $g(t)$, such that
 $g(T_0) \sim F_0^*$ and $g(T_1) \sim F_1^*$, where
$F_0^*$ and $F_1^*$ are exponential distributions, then $F_0$ and  $F_1$ have proportional hazards
\citep[see e.g.,][Supplement]{fay2018confidence}. We emphasize that we only need to know that $g(\cdot)$ exisits, we do not need to know what it is.
\end{enumerate}
Thus, by condition~\ref{PH:increasing.Transformation.to.Exponential} we see that
the proportional hazards model is more general than the two exponentials model.

A mechanistic motivation for the proportional hazards model is to assume that the instantaneous event rate (i.e., the hazard rate) for the control arm can change
over time and is equal to $\lambda_0(t)$, and we have a leaky vaccine, meaning that it acts the same on all individuals by changing the
hazard rate at every time point by a constant regardless of the time since vaccination. This gives $\lambda_1(t)= \theta_{Cox} \lambda_0(t)$.
Again we assume a homogeneous population at risk for every time point
(see Section~\ref{sec-heterogeneity} for violations of this assumption).




\subsection{Cumulative Hazard}
\label{sec-CH}

The estimand
 $\theta_{Cox}(\tau)$
is like a ratio of weighted averages of the hazards
(equation~\ref{eq:VE.Cox.weighted.mean.of.hazards}), where
the weights are larger earlier on in the study (equation~\ref{eq:weights.for.Cox}).
The larger weighting earlier in the study makes some sense, because $\lambda_a(t)$ represents the instantaneous probability of an event at $t$ given survival until just before $t$, and the proportion surviving, $S_a(t)$, is  larger earlier in the study.

In this section, we motivate the cumulative hazard ratio, say $\theta_{CH}(\tau)$, as a reasonable
weighted average of local hazard ratios. Under proportional hazards, any weighted average
of hazard ratios will equal $\theta_{CH}(\tau) = \theta_{Cox}(\tau)$, so this motivation is about
finding an alternative to $\theta_{Cox}(\tau)$ that equals the usual hazard ratio under proportional hazards, but has a more intuitive interpretation when the proportional hazards assumption fails.
The value $\theta_{CH}(\tau)$ or some closely related measures have been suggested in the survival literature as a more robust estimand alternative to $\theta_{Cox}(\tau)$ \citep{wei2008estimating,vansteelandt2024assumption}.
 Because the cumulative hazard ratio is
rarely, if ever, used to create a VE estimand, we start with its motivation as a weighted average of local hazard ratios.

Let a weighted mean of hazard ratios over the course of the study be,
\begin{eqnarray}
\theta_{wh}(\tau) & = & \frac{ \int_0^{\tau} w(t) \theta_h(t) dt }{ \int_0^{\tau} w(t) dt },
\label{eq:theta.whr}
\end{eqnarray}
where $\theta_h(t) = \lambda_1(t)/\lambda_0(t)$ is the (local) hazard ratio at time $t$, and $w(t)$ is a  weight function \citep[see e.g.,][]{kalbfleisch1981estimation,schemper2009estimation}.
When $w(t) \propto \lambda_0(t)$ then $\theta_{wh}(\tau)=\theta_{CH}(\tau) = \frac{\Lambda_1(t)}{\Lambda_0(t)},$
where $\Lambda_z(t) = \int_0^t \lambda_z(u) du$ is the cumulative hazard for arm $z$ up to time $t$.

Consider a mechanistic motivation for $w(t) \propto \lambda_0(t)$.
Let  $\epsilon(t)$ be the average exposure rate at time $t$ in the study.
 By blindedness and randomization we assume that
$\epsilon(t)$ is the same in both arms.
For simplicity we have assumed that all individuals in the trial started at the same time, so that the time since randomization will be the same calendar time for all individuals.
For staggered accrual, we can use the same model, but the interpretation of $\epsilon(t)$ requires
more nuance. In that case, $\epsilon(t)$ represents an average exposure effect based on
all individuals that got randomized $t$ time units earlier, and this average exposure effect will span over many calendar time points because of the differing randomization times within the study.
In this section, we assume that there is no change over time in the nature of the population still at risk of getting the event for the first time. This assumption would be violated if there are different types of individuals,
such that some types are more likely to get the event earlier due to either some aspect of their lifestyle that affects exposure or some aspect of their immune system  (see Section~\ref{sec-heterogeneity} for violations of that assumption).
Let $\phi_0$ be the rate of events given exposed  in the control population, which does not depend on
$t$, the time since vaccination, because we assume that the control vaccine does not affect the
event of interest.
Then the hazard of an event in the control population is
\begin{eqnarray*}
\lambda_0(t) & = & \epsilon(t) \phi_0.
\end{eqnarray*}
For the vaccinated population that were vaccinated
at time $0$, let $\phi_1(t)$ be the rate of events given exposed at the average exposure rate, $\epsilon(t)$. Unlike the control population, the effect of the vaccinated population conditioned on exposure changes based on time since vaccination.
Then,
\begin{eqnarray*}
\lambda_1(t) & = & \epsilon(t) \phi_1(t).
\end{eqnarray*}
The (local) hazard ratio effect of the vaccination at time $t$  is
\begin{eqnarray}
\theta_h(t) & = &   \frac{ \lambda_1(t) }{ \lambda_0(t) } = \frac{ \epsilon(t) \phi_1(t) }{ \epsilon(t) \phi_0 }  = \frac{  \phi_1(t) }{ \phi_0 },
\label{eq:theta.h.eq.phi1.over.phi0}
\end{eqnarray}
and $\theta_h(t)$ does not depend on  $\epsilon(t)$.

Now suppose we want to summarize the vaccine effect from time $t=0$ to time $t=\tau$
using $\theta_{wh}(\tau)$, a weighted average of $\theta_h(t)$ over time.
Information about vaccine efficacy  is related to exposure. We can see this in the limit, since if $\epsilon(t)=0$, then no one will be exposed at time $t$ and we have no information on whether the vaccine works or not during that time period.
Thus, it is reasonable to  have the weight function be proportional to $\epsilon(t)$, since the larger the exposure, the more information we have about whether the vaccine works.
Further, since $\phi_0$ does not depend on $t$, we have $\lambda_0(t) \propto \epsilon(t)$. Since
$\lambda_0(t)$ is identifiable, we can use that as a weight, so to get a summarization of the vaccine effect from time $t=0$ to time $t=\tau$, we could use
\begin{eqnarray*}
 \frac{ \int_0^{\tau} \lambda_0(t) \theta_h(t) dt }{ \int_0^{\tau} \lambda_0(t)  dt} & = & \frac{  \Lambda_1(\tau)  }{ \Lambda_0(\tau) } = \theta_{CH}(\tau).
\end{eqnarray*}
By equation~\ref{eq:theta.h.eq.phi1.over.phi0}, $\theta_h(t)$ does not depend on the exposure rate (as long as it is greater than 0), and  the effect of exposure on $\theta_{CH}(\tau)$ is only through the weighting function.
The VE associated estimand is
\begin{eqnarray}
VE_{CH}(\tau) & = & 1 - \frac{ \Lambda_1(\tau) }{ \Lambda_0(\tau) } = 1-\theta_{CH}(\tau), \label{eq:VE.CH}
\end{eqnarray}
In continuous time, we have that $S_z(t) = \exp( - \Lambda_z(t) )$ \citep[see e.g.,][]{aalen2008survival}, so we can rewrite $VE_{CH}(\tau)$ as
\begin{eqnarray}
VE_{CH}(\tau) & = & 1 - \frac{ \log \left\{ S_1(\tau) \right\} }{ \log \left\{  S_0(\tau) \right\} }. \label{eq:VE.CH.using.logS.ratios}
\end{eqnarray}

Since the interpretation of a ratio of log survival function is not intuitive,
we  use algebra on the log survival ratio to get
$S_1(\tau)= S_0(\tau)^{\theta_{CH}(\tau)}$. Here is an example interpretation.
If $VE_{CH}(\tau)=70\%$ then by the end of the study,  the expected proportion without an event in the test vaccine arm is equal to the expected proportion without an event in the control vaccine arm
raised to the power $0.30$ (e.g., if 90\% do not have an event in the placebo arm, then $0.90^{0.30}=96.9\%$
in the vaccine arm will not have an event).
Even this is not a simple lay explanation.

For another explanation, first, call a hazard function a {\it force of infection} \citep[see e.g.,][p. 26]{halloran2010design}. Then a $VE_{CH}(\tau)=70\%$ results in
a 70\% reduction in the weighted average force of infection in the vaccinated population compared to in the unvaccinated population, where the weight at any time is proportional to the exposure rate in the unvaccinated population at that time.

A constancy model is the proportional hazards model, which was previously discussed at the end of  Section~\ref{sec-cox}, and all of those interpretations under proportional hazards hold.

All of the estimands in Section~\ref{sec-cumulative} (including Section~\ref{sec-odds}) are defined in continuous time, but in practice data are collected in discrete time.  This only makes a difference for the estimands that are based on hazards. For completeness, we discuss the weighted average discrete hazard in Appendix~\ref{sec-weighted.avg.discrete.haz} which shows the differences between a discrete and a continuous hazard; however, the differences are slight and the weighted average discrete hazard
estimand does not lead to much insight and will not be discussed further.

\subsection{Odds}
\label{sec-odds}

Sometimes the odds ratio is suggested for a ratio
estimator and 
estimand.
The odds ratio estimand
is used primarily in two situations. First,  as an approximation for rare diseases (when $F_0(\tau)$ is small) to one of the others mentioned \citep{halloran1997study}. Second, for use in case-control studies or the closely related test-negative study designs \citep[see e.g.,][]{sullivan2016theoretical,westreich2016invited}.
The odds ratio estimand in the format and context of this paper is
\begin{eqnarray}
\theta_{odds}(\tau) & = &  \frac{ \frac{F_1(\tau)}{1-F_1(\tau)} }{ \frac{F_0(\tau)}{1-F_0(\tau)} }  \\
& = &
\frac{F_1(\tau)}{ F_0(\tau) } \left( \frac{ 1- F_0(\tau) }{ 1 - F_1(\tau) } \right) \nonumber
\end{eqnarray}



The simple lay explanation for $VE_{odds}(\tau)=70\%$  is that
the odds of an event by time $\tau$ after vaccination is reduced by 70\% in a  vaccinated population compared to an unvaccinated population.

The constancy model is the proportional odds model, where $\theta_{odds}(t)=\theta_{odds}$ is a constant
that does not depend on $t$.

We know of no mechanistic model to motivate the odds ratio. The odds ratio is mostly used
as a mathematical convenience because it can be estimated consistently in case-control studies
without knowing the control incidence rate for the study, $F_0(\tau)$,
and odds ratios are an easily parameterized effect after making covariate adjustments through logistic regression \citep[see e.g.,][]{breslow1996statistics}.

\section{Comparing ITT Cumulative VE Estimands}
\label{sec-comparison.of.estimands.general.case}

First consider the case with low event rates, such that $F_0(\tau) \approx 0$ and $F_1(\tau) \approx 0$.
Then since $S_0(\tau) \approx 1 \approx S_1(\tau)$,  then $VE_{odds}(\tau) \approx VE_{CI}(\tau)$. Additionally,  $\mu_0(\tau) \approx \mu_1(\tau)$ so
$VE_{CI}(\tau) \approx VE_{IR}(\tau)$. Under the low event rate assumption, $w(t;\theta) \approx 1/(\theta+1)$ in the definition of
$VE_{Cox}$ so that $VE_{Cox}(\tau) \approx VE_{CH}(\tau)$.
Finally, since $S_z(\tau) \approx 1$,  then $-\log(S_z(\tau) ) \approx F_z(\tau)$ and
$VE_{CH}(\tau) \approx VE_{CI}$. Thus, with low event rates
\begin{eqnarray*}
VE_{CI}(\tau) & \approx & VE_{IR}(\tau)  \approx
VE_{Cox}(\tau) \approx VE_{CH}(\tau) \\
& \approx &  VE_{odds}(\tau).
\end{eqnarray*}
To operationalize how low the event rates need to be to give good approximations, we give more details.

\begin{figure}[tb!]
\includegraphics[width=6in]{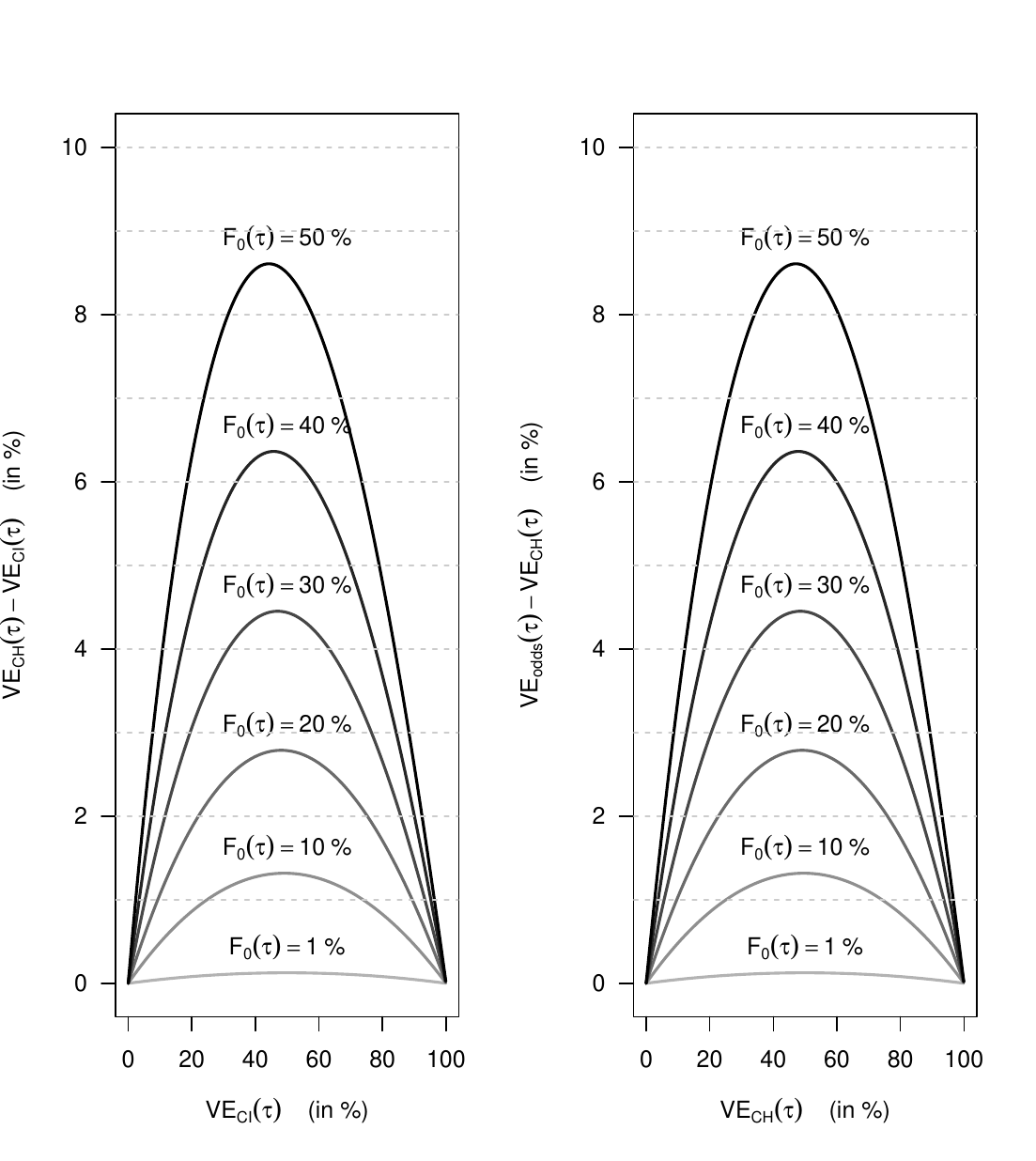}
\caption{\tiny Comparison of $VE_{CH}(\tau) - VE_{CI}(\tau)$ in percent versus  $VE_{CI}(\tau)$ in percent (left panel) and $VE_{odds}(\tau) - VE_{CH}(\tau)$ in percent versus  $VE_{CH}(\tau)$ in percent (right panel). Different lines represent different values of $F_0(\tau)$. Maximum differences for both panels are
0.13\% (when $F_0(\tau)=1\%$),
1.32\% (when $F_0(\tau)=10\%$),
2.79\% (when $F_0(\tau)=20\%$),
4.45\% (when $F_0(\tau)=30\%$),
6.36\% (when $F_0(\tau)=40\%$), and
8.61\% (when $F_0(\tau)=50\%$).
 \label{fig:VE2plots}
 }
\end{figure}

The estimands $VE_{CI}(\tau)$, $VE_{CH}(\tau)$ and $VE_{odds}(\tau)$ are defined by only $F_0(\tau)$ and $F_1(\tau)$; therefore, we can relate them to each other algebraically given $F_0(\tau)$.
It is easier to write the relationships in terms of ratio effects.
We can transform $VE_{CI}(\tau)=1-\theta_{CI}(\tau)$ into $VE_{CH}(\tau)=1-\theta_{CH}(\tau)$ using the following expression
\begin{eqnarray}
\theta_{CH}(\tau) & = & \frac{ \log \left\{ 1-  \theta_{CI}(\tau) F_0(\tau)  \right\} }{ \log \left\{  1-F_0(\tau) \right\} }. \label{eq:theta.CI.into.theta.CH}
\end{eqnarray}
Rewriting $\theta_{CI}(\tau)$ in terms of $\theta_{odds}(\tau)$ and $F_0(\tau)$ is analogous to a standard
formula for converting  odds ratios to risk ratios  \citep[see e.g.,][]{zhang1998s},
\begin{eqnarray}
\theta_{CI}(\tau) & = & \frac{\theta_{odds}(\tau)}{1-F_0(\tau) + \theta_{odds}(\tau) F_0(\tau)}
\label{eq:thetaCI.using.thetaOdds}
\end{eqnarray}
If the vaccine works then $F_1(\tau) < F_0(\tau)$ and we can show using the relationship between VEs
(i.e., using equations~\ref{eq:theta.CI.into.theta.CH} and \ref{eq:thetaCI.using.thetaOdds}) that
 $VE_{CI}(\tau) < VE_{CH}(\tau) < VE_{odds}(\tau).$
To see the practical importance of
those inequalities, we plot VE differences in Figure~\ref{fig:VE2plots}.
In the left panel we plot $VE_{CI}(\tau)$ by $VE_{CH}(\tau) - VE_{CI}(\tau)$ in terms of percent
for different values of $F_0(\tau)$, while in the right panel we plot $VE_{CH}(\tau)$ by $VE_{odds}(\tau) - VE_{CH}(\tau)$.
Remarkably, the peak difference for any fixed value of $F_0(\tau)$ is the same for both panels (see Appendix~\ref{app-peakVEdiff.equivalence}).
Thus, $VE_{CH}(\tau)$ is approximately halfway between $VE_{CI}(\tau)$ and $VE_{odds}(\tau).$

Other estimands such as $VE_{IR}(\tau)$ and $VE_{Cox}(\tau)$ depend on the values of $F_0(t)$ and $F_1(t)$ for $t < \tau$.
Consider first,  $\theta_{IR}(\tau) = \frac{ F_1(\tau)/\mu_1(\tau)}{ F_0(\tau)/\mu_0(\tau)} = \theta_{CI}(\tau) \frac{ \mu_0(\tau)}{ \mu_1(\tau)}$.
By the definition of $\mu_z(\tau)$ and the monotonicity of $S_z(t)$, we have that
\begin{eqnarray*}
S_z(\tau) \tau \leq \mu_z(\tau) = \int_0^{\tau} S_z(t) dt \leq \tau.
\end{eqnarray*}
So
\begin{eqnarray*}
\frac{\tau S_0(\tau)  }{\tau} \leq  \frac{ \mu_0(\tau)}{\mu_1(\tau)}  \leq \frac{\tau}{\tau S_1(\tau)}
\end{eqnarray*}
and
\begin{eqnarray}
\theta_{CI}(\tau) \left\{ 1- F_0(\tau) \right\}  \leq \theta_{IR}(\tau) \leq
\frac{ \theta_{odds}(\tau) }{  1- F_0(\tau)   }.
\label{eq:thetaIR.inequalities}
\end{eqnarray}
Now consider $\theta_{Cox}(\tau)$,
which weighs earlier hazard ratios larger than later hazard ratios.
As one might expect from the definition of $\theta_{Cox}(\tau)$ (see
equation~\ref{eq:VE.Cox.weighted.mean.of.hazards}),
it is difficult to make simple inequality statements about it.

\begin{figure}[bt]
\includegraphics[width=6in]{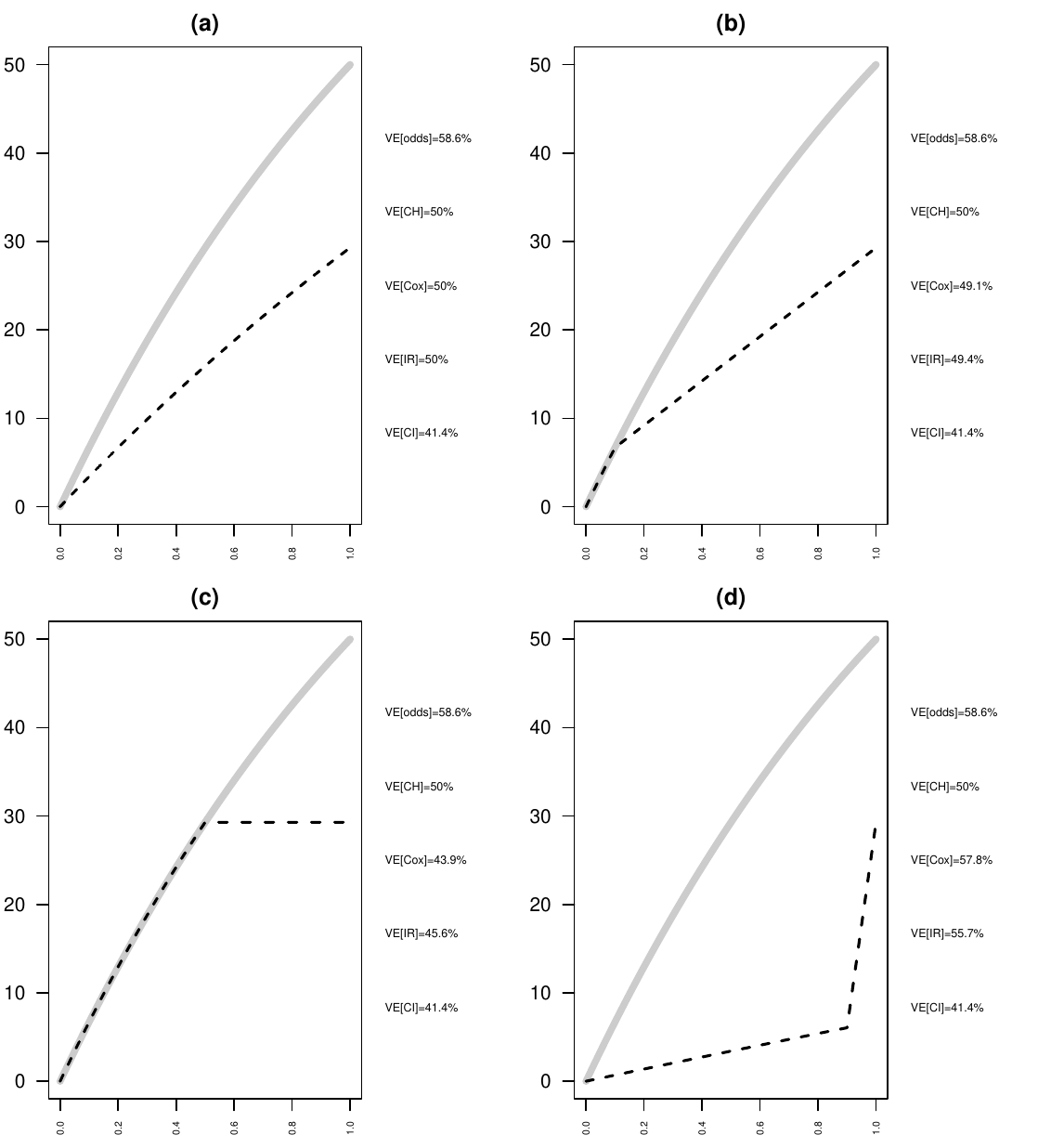}
\caption{Examples of cumulative VE estimands (right text for each panel), horizontal axes are $t$ and vertical axes cumulative incidence (in \%), with $F_0(t)$ (gray, solid) and $F_1(t)$ (black, dashed). Panel~(a):Two exponentials with  $\lambda_0= -\log(-0.5)$ so that $F_0(1)=50\%$,
and $\lambda_1= 0.50 \lambda_0$.
Panel~(b): Same as (a), except for $t<0.1$ set $F_1(t)=F_0(t)$, and for $t \geq 0.1$ linearly extrapolate from $F_1(0.1)$ to  $F_1(1)=50\%$.
Panel~(c): Same as (a), except for $t<0.5$ set $F_1(t)=F_0(t)$, and for $t \geq 0.5$ linearly extrapolate from $F_1(0.5)$ to  $F_1(1)=50\%$.
Panel~(d): same as (a), except for $t<0.9$ let $F_1(t)$ be an exponential with rate $\lambda_1=\lambda_0/10$, giving an early $VE_{CH}(0.9)=90$\%, and  for $t \geq 0.9$ linearly extrapolate from $F_1(0.9)$ to  $F_1(1)=50\%$.
\label{fig:plotVEs.a.to.d}
 }
\end{figure}

In Figure~\ref{fig:plotVEs.a.to.d} we give a few specific examples to graphically compare VE estimands when $\tau=1$.
Figure~\ref{fig:VE2plots} has already shown very generally differences between $VE_{CI}(\tau)$,
$VE_{CH}(\tau)$ and $VE_{odds}(\tau)$, so in Figure~\ref{fig:plotVEs.a.to.d} we emphasize $VE_{Cox}(\tau)$
and $VE_{IR}(\tau)$ in the $F_0(\tau)=0.50$ case, where
there is a large difference between $VE_{CI}(\tau)$ and $VE_{odds}(\tau)$.
All 4 panels of Figure~\ref{fig:plotVEs.a.to.d} have the same values for $F_0(\tau)$ and $F_1(\tau)$
so the values of $VE_{CI}(\tau)$,
$VE_{CH}(\tau)$ and $VE_{odds}(\tau)$ are the same for all 4 panels.
In Panel~(a) we show the two exponential case, where theoretically
$VE_{CH}(\tau) = VE_{Cox}(\tau) = VE_{IR}(\tau)$.
In Panel~(b) we force $F_0(t)=F_1(t)$ to be equal for $t<0.1$, then linearly interpolate to the
same values of $F_0(1)$ and $F_1(1)$ as in Panel~(a). We see that $VE_{Cox}(\tau)$ and $VE_{IR}(\tau)$
are approximately equal to $VE_{CH}(\tau)$.
Panels~(c) and (d) are extreme cases. Panel~(c) is similar to Panel~(b) except we force
$F_0(t)=F_1(t)$ to be equal for $t<0.5$, and in this case $VE_{Cox}(\tau)$ and $VE_{IR}(\tau)$
are much closer to $VE_{CI}(\tau)$ than $VE_{odds}(\tau)$.
Panel~(d) is an extreme case in the opposite direction
 where there is a very strong hazard ratio effect early on,
but it drops off drastically after $t>0.9$, and in that case
$VE_{Cox}(\tau)$ and $VE_{IR}(\tau)$
are much closer to $VE_{odds}(\tau)$ than $VE_{CI}(\tau)$.
In all the panels
the $VE_{IR}(\tau)$ follows closely with $VE_{Cox}(\tau)$.  This can change because $VE_{Cox}$ is rank-based while $VE_{IR}$ is based also on restricted mean survival times.



\section{Ramp-Up Models}
\label{sec-ramp-up}

In the previous sections, we have assumed that the population of interest is the ITT population, all randomized individuals in the study (or the superpopulation associated with them).
However, in order to measure the cumulative full immunization effect  of the vaccine (often called the  `biological effect' \citep{horne2001intent}), some individuals in the study will not give the vaccine a fair chance to show how it would work when used as directed.
These are individuals that have one or both of the following conditions: (1) they would not adhere to at least one of the vaccine regimens if assigned to it, and/or (2) they would have the event before time $t_{RU}$ in at least one of the vaccine regimens if assigned to it, where $t_{RU}$ is the amount of time since full vaccination before the vaccine has had a chance to `ramp-up' to its full efficacy (e.g., by having enough time to create antibodies against the agent of interest).
Ideally, if we could identify those individuals and exclude them just prior to randomization, that
would define a full immunization effect on the rest of the population, say the adherent ramp-up period surviving (A-RUPS) population.
Measuring any of the cumulative VE estimands on the A-RUPS population
defines a full immunization VE estimand.
In that case, the
A-RUPS population  acts as a principle stratum for the purposes of defining a causal estimand \citep{frangakis2002principal}.
\citet{gilbert2013sensitivity} describe these types of full immunization causal effects, calling one of them    ``survivor causal effect[s]'' that account for adherence.
They detail the assumptions necessary to identify these causal effects which may be applied to an individual randomized trial.


To simplify and focus this paper, we assume
 that we can identify
the population that would adhere to both arms of the study, and we can use that population to
define the distributions of the two arms of the study.
 Then in the following,
we define the ramp-up versions of the VE estimands that account for only measuring events
after the predetermined ramp-up period.
Alternatively, we can define the ramp-up version of the VE estimands on the full population,
this will give full {\it practical} immunization effects,
 where the `practical' indicates a real-world effect since
some people do not get (perhaps cannot tolerate) the recommended vaccine course.
The full practical immunization effects require many fewer assumptions to identify,
because the adherence population is defined as adhering to whichever arm one is randomized to, but
in practice we only observe adherence to the actual arm that participants are in.
Since our presentation will focus on the ramp-up period, we call either type of VE estimand
a
 {\it ramp-up } estimand. In the following, we discuss only the full practical immunization effects,
so we do not have to introduce new notation for the adherent population, but the discourse easily applies to the full immunization effect by switching to the adherent population.

An impractical study design to capture full practical immunization effects would be to randomize then
isolate all randomized participants from exposure to the infectious agent
during their ramp-up period.
Because this design is rarely feasible, we consider only the usual randomized trial where participants are naturally exposed throughout the study.

We address allowing the necessary time for the vaccine to work by a ramp-up model.
Define an ideal ramp-up model as any $F_0(t)$ and $F_1(t)$ such that $F_0(t)=F_1(t)$ for all $t \leq t_{RU}$. An ideal ramp-up model would hold for a test vaccine that acts the same (with respect to the occurrence of events) as the control vaccine up until $t_{RU}$. This is an overly simplified model of reality, but it is a step in the direction of measuring the full practical immunization effect of the vaccine.

Under an ideal ramp-up model, we create VE estimands by comparing conditional distributions.
Let
\begin{eqnarray*}
Pr [ T(z) \leq t | T(z) > t_{RU} ] & = &   \frac{F_z(t) - F_z(t_{RU})}{1-F_z(t_{RU}) }, \\
& & \hspace*{4em} \mbox{ when $t>t_{RU}$, }
\end{eqnarray*}
for $z=0,1$. Let $t^*= t - t_{RU}$, which represents the time since the end of the ramp-up period.
Then define the cumulative distribution conditioning on no event before $t_{RU}$ as
\begin{eqnarray}
F^*_z(t^*) & = &   \frac{F_z(t_{RU} + t^*) - F_z(t_{RU})}{1-F_z(t_{RU}) }, \\
& & \hspace*{4em}  \mbox{ when $t* > 0$.} \nonumber
\label{eq:Fstar.defn}
\end{eqnarray}
Then we can replace $F_0(t)$ and   $F_1(t)$ for $t>0$, with
  $F_0^*(t^*)$ and   $F_1^*(t^*)$ for $t^*>0$ in any of the previous ITT VE estimands, to get a
ramp-up version of the VE estimand. We will call this a ramp-up VE estimand even if the {\it ideal}
ramp-up model does not hold (i.e., even if $F_0(t) \neq F_1(t)$ for some $t \leq t_{RU}$).
One  example is $VE^*_{CI}(t^*) = 1 - \frac{F_1^*(t^*)}{F_0^*(t^*)}$, which
gives a ramp-up version of the cumulative incidence VE estimand.
\citet{michiels2022estimation} in their Section C.1 show (using different terminology and notation) that
under the ideal ramp-up model, when $t> t_{RU}$, then
\begin{eqnarray}
VE_{CI}^*(t^*) & = & VE_{CI}(t) \left( \frac{F_0(t) }{F_0(t)- F_0(t_{RU})} \right), \label{eq:VECI.VECIstar}  \\
& &  \hspace*{4em} \mbox{  where $t^*=t-t_{RU}$.}  \nonumber
\end{eqnarray}
This does not hold if we cannot assume the ideal ramp-up model. Unfortunately, even for ideal ramp-up models such simplifications  are not straightforward for ramp-up versions of the other VE estimands studied in Section~\ref{sec-cumulative}.

Equation~\ref{eq:VECI.VECIstar} shows how if the ideal ramp-up model holds, then if $F_0(t_{RU})=F_1(t_{RU})$ is not small compared to $F_0(t)$, then there can be a substantial difference between the ramp-up VE estimand ($VE_{CI}^*(t^*)$) and the intention-to-treat estimand
($VE_{CI}(t)$).

Without assuming the ideal ramp-up model,  the ramp-up VE estimands are still well defined
as functionals of the distributions of potential outcomes;
 however,
their meaning is not necessarily straightforward.
The issue is that the conditional populations are different when the ideal ramp-up model fails.
Let $\Omega_z = \left\{ i: T_i(z) > t_{RU} \right\}$ represent a conditional population for arm $z$.
Then even under a fair randomization, when the ideal ramp-up model fails (or even if $F_0(t_{RU})=F_1(t_{RU})$  but the individuals that fail before $t_{RU}$ are different in the two arms), then
$\Omega_0 \neq \Omega_1$, and comparing $F^*_0$ to $F_1^*$ are comparing different populations,
and hence do not have a clear causal interpretation.

\begin{figure}[bt]
\includegraphics[width=6in]{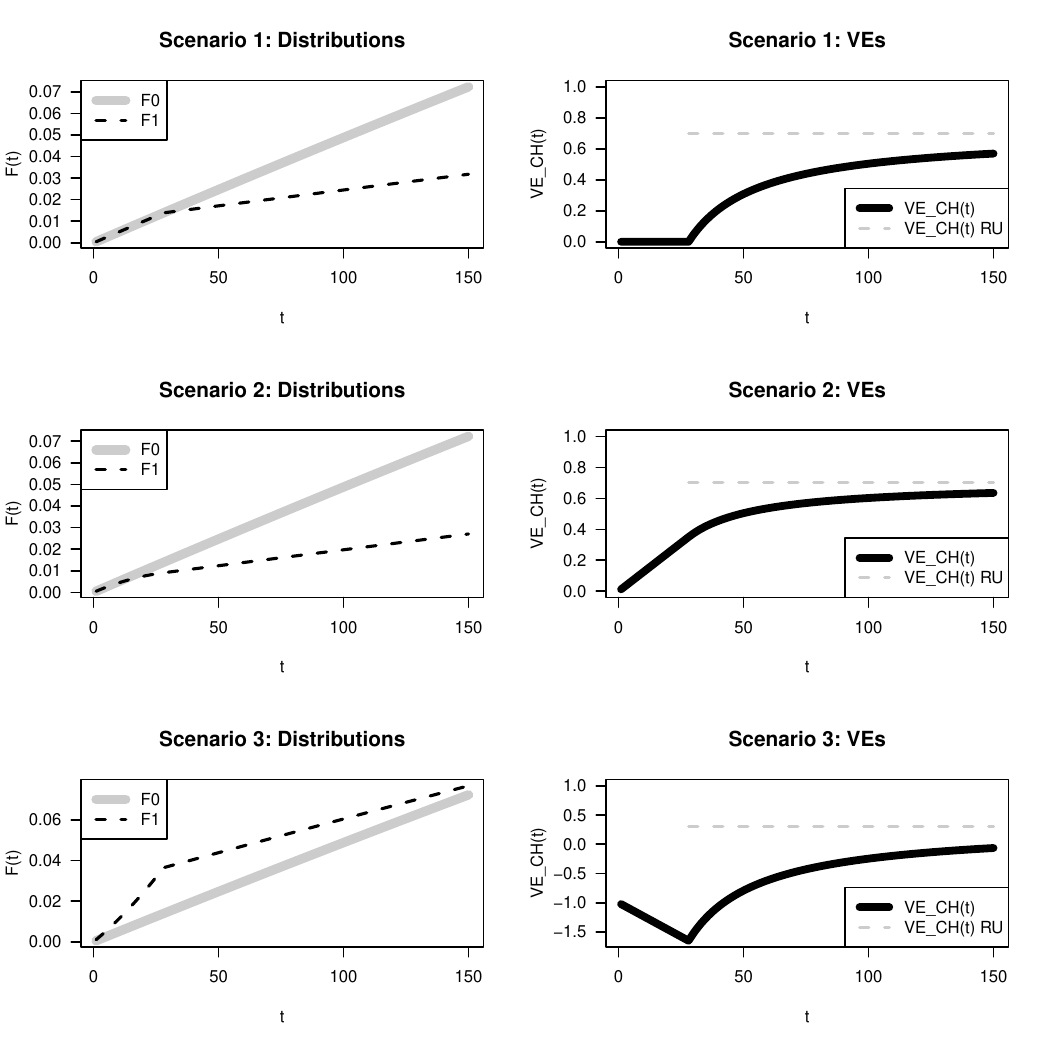}
\caption{$VE_{CH}(t)$ and $VE_{CH}^*(t^*)$ (labeled as 'VE\_CH(t) RU') in three scenarios.
Since $t^*=t-t_{RU}$,  $VE_{CH}^*(t^*)$ is defined only for $t^*>0$ or equivalently $t>t_{RU}$.
Loosely, the 3 scenarios are: (1) equal distributions during ramp-up period,
then vaccine instantly works full force with local hazard ratio equal to  $\theta_h=0.30$,
(2) same as scenario 1, but the vaccine gradually begins to work during the ramp-up period,
(3) the test vaccine harms ($\theta_h(0)=2$) at the beginning of the ramp-up period  then gradually gets to local hazard ratio  that helps ($\theta_h=0.70$) by the end of the ramp-up period and remains there. (Details in Appendix~\ref{app:RampUpCumHaz}).
\label{fig:RampUpCumHaz}
 }
\end{figure}

To see the difficulties, consider the ramp-up VE estimand based on the
cumulative hazard, specifically,
\begin{eqnarray*}
VE_{CH}^*(t^*) & = & 1 - \frac{ \log \left\{ 1- F_1^*(t^*) \right\} }{\log \left\{ 1- F_0^*(t^*) \right\}},
\end{eqnarray*}
where $F_z^*(t^*)$ for $z=0,1$ is defined in equation~\ref{eq:Fstar.defn}.
From Section~\ref{sec-CH} we know that if $F_1^*(t^*)$ and $F_0^*(t^*)$ relate by proportional hazards,
then  $VE_{CH}^*(t^*)$ is a constant representing the ratio of local hazards for $t^*>0$ (i.e., $t > t_{RU}$).
In Figure~\ref{fig:RampUpCumHaz} we compare $VE_{CH}(t)$ and $VE_{CH}^*(t^*)$ in three such scenarios.
In all three scenarios $VE_{CH}^*(t^*)$ for $t^*>0$ is equal to a constant $>0$, but they differ prior to $t_{RU}$.
Details of the scenarios are given in Appendix~\ref{app:RampUpCumHaz}.
Scenario~1 is an ideal ramp-up model ($F_0(t)=F_1(t)$ for $t<t_{RU}$) then $VE_{CH}^*(t^*)=0.70$. This model is not biologically realistic because the test vaccine effect goes from no effect to its full effect instantaneously at $t_{RU}$.
Nevertheless, we see that in the ideal situation, $VE_{CH}^*(t^*)=0.70$ captures the full immunization effect nicely. Scenario~2 is more realistic; it is the same as Scenario~1 except that the
effect of the test vaccine  slowly builds up during the ramp-up period until it is at full force at $t_{RU}$ and $VE_{CH}^*(t^*)=0.70$ for all $t^*>0$. Notice that $VE_{CH}(t)$ captures this.
Scenario~3 is a cautionary example, where the test vaccine enhances the probability of the event at the beginning of the ramp-up period ($\lim_{t \rightarrow 0^+} \theta_h(t)=2$), then the local hazard gradually moves toward a local hazard ratio of
$\theta_h(t_{RU})=0.70$ at $t_{RU}$ and remains there for $t>t_{RU}$.
In this scenario, we see that the ramp-up VE estimand $VE_{CH}^*(t^*)=0.30$ for $t^*>0$ apparently shows a substantial benefit of the test vaccine after $t_{RU}$, even though by the end of the study more individuals are expected to have the event in the test vaccine arm than in the control arm (see  the left panel showing the distributions $F_0$ and $F_1$).
The VE estimand $VE_{CH}(t)$ captures the harm of the test vaccine in this scenario, since its
values are all less than $0$. Scenario~3 highlights that even though we have a fair randomization
and can compare functions of $F_0$ and $F_1$, when we compare those functions by creating the conditional effects $F_0^*$ and $F_1^*$, they can be misleading about the usefulness of a vaccine. In practice, it is wise to always plot the incidence curves, so that departures from the ideal ramp-up model are readily apparent.


\citet{michiels2022estimation} consider a different estimand to handle the ramp-up period.
They discuss vaccine efficacy if the ramp-up period could be eliminated
and use a method ``inspired by structural accelerated failure time models'' using a structural distribution model.  Their model allows the effect during the ramp-up period to be multiplied by a factor between $0$ (no vaccine effect during ramp-up period) and $1$ (vaccine effect is the same during ramp-up period and after it).

\section{Heterogeneous Risk Models and Changing Hazard Rates Over Time}
\label{sec-heterogeneity}



This section is about local effects, when we want to study VE estimands over time to see how the
vaccine effect is changing over time.
Although the cumulative VE estimands depend on $\tau$ (the last time point in the study), we cannot just replace $\tau$ with $t$ in those estimands, because then it would measure the cumulative effect up to time $t$, not the local effect.
Typically, local effects are studied by plotting the VE estimand over time using local hazard ratios, $VE_{h}(t) = 1- \theta_h(t) = 1 - \frac{\lambda_1(t)}{\lambda_0(t)}$.
This appears to make sense, because for each study arm the hazard at $t$ is the instantaneous rate of an event occurring at $t$ conditioning on not having the event before $t$.
The problem is when we incorrectly impose an individual-level causal interpretation of the hazard ratio over time, meaning that $\theta_h(t)$ represents on average the change in the hazard of an individual if vaccinated
compared to themselves if on control.
Recall in the description of the mechanistic derivation of the proportional hazards model that we
made not only the proportional hazards assumption but also
assumed that the population was homogeneous in the sense that the nature of the population was not changing over time.  In this section, we discuss violations of both assumptions.

\citet{lin2022reliably} addresses the violation of the proportional hazards assumption only.
Often researchers will address non-proportional hazards by simple models that assume constant
$VE$ estimators within each of several successive time periods using a time varying effect in a Cox model
or a Poisson model.  \citet{lin2022reliably} suggest a more sophisticated approach that accounts for
staggered accrual in the study.
They use a Cox model with two time indexes,  modeling the calendar time effect semi-parametrically
and modeling the time since vaccination  using a piecewise linear model for the log hazard ratio.
\citet{lin2022reliably} show graphically on some simulated data how their method is more precise
at estimating $VE_{h}(t)$ than the simple piecewise constant hazard model.

An issue is that no matter how well you estimate $VE_h(t)$, in order to interpret
$VE_h(t)$ as representing the average individual-level effect of the vaccine (e.g., whether the
vaccine is waning in its effect as the time since vaccination increases), one needs to assume a homogeneous study population over time.
The violation of that assumption can cause depletion of susceptibles bias (see e.g., \citet{hernan2010hazards} or  \citet{fay2024causal} and its references).
It is easiest to explain this issue using the concept of frailty,
which is reviewed nicely in  \citet{aalen2008survival}, Chapter 6, and which we discuss next.

Consider a data generating mechanism whereby the hazard for the potential outcome    $T_i(0)$ for individual $i$ is
$\lambda_0^{id}(t; U_i) = U_i \epsilon(t) \phi_0$, where $U_i$ is a positive factor known as the frailty, and as in Section~\ref{sec-CH} $\epsilon(t)$ represents the
average exposure rate of an individual vaccinated $t$ time units earlier in the study population, and $\phi_0$ is a factor to translate exposure rate to a reference hazard rate  (i.e., the hazard rate when $U_i=1$) for the control potential outcome.
For the test vaccine potential outcome for individual $i$, we use
 $\lambda_1^{id}(t; U_i) = U_i \epsilon(t) \phi_1(t)$, where we allow the multiplicative  effect of the
 vaccine to change based on $t$.
We use the ``id'' superscript to differentiate this individual hazard function from the population hazard function we have used previously.
In this model,  $U_i$ is associated with the $i$th individual and the factor is the same for both potential outcomes. This induces correlation among the potential outcomes.
The frailty is a catchall multiplicative factor encompassing all aspects of individual $i$ that make that individual more or less likely to have an event at that time than a modelled reference person.
For example, higher values of $U_i$ could mean that individual $i$ is more likely to be exposed to the agent of interest than a person with lower frailty and/or has a better non-specific immune system (acting independent of vaccination) than a reference person.
Under this model, the {\it individual} hazard ratio effect for the $i$th individual is
\begin{eqnarray*}
\theta^{id}_h(t) & = & \frac{ \lambda_1^{id}(t; U_i)  }{ \lambda_0^{id}(t; U_i) } = \frac{ U_i \epsilon(t) \phi_1(t) }{ U_i \epsilon(t) \phi_0 } = \frac{ \phi_1(t) }{ \phi_0 }
\end{eqnarray*}
and  all individual's have the same individual hazard ratio effect because the frailty effects cancel out.
Further, $\theta^{id}_h(t)$ does not depend on the average exposure effect, $\epsilon(t)$.

Although $\theta^{id}_h(t)$ has a nice mechanistic interpretation, it is not identifiable from an individual randomized trial without baseline covariates. If there are baseline covariates that are highly correlated with the frailty, then building a model using those covariates (e.g., a Cox model with covariates) may give estimators consistent for estimands closer to the individual level estimands \citep[see e.g.,][]{fay2024causal}.
Without those covariates, the time dependent population hazard ratio,
$\theta_h(t) =  \lambda_1(t)/\lambda_0(t)$, depends on the frailty and exposure distribution.

Suppose the $U_i$ are independently distributed gamma with a mean 1 and variance $\nu$.
Then it can be shown that the population hazards for arm $z$ in our model are \citep[see e.g.,][equation~6.8]{aalen2008survival},
\begin{eqnarray*}
\lambda_z(t) & = & \frac{ \lambda_z^{id}(t;1) }{1+ \nu \int_0^{t} \lambda_z^{id}(u;1) du}
 =  \frac{ \lambda_z^{id}(t;1) }{1-  \nu \log \left\{ S_z^{id}(t;1) \right\} }
\end{eqnarray*}
where $S_z^{id}(t;1)=1-F_z^{id}(t;1)$ is the survival distribution for an individual with $Z_i=z$ and $U_i=1$.
Then the population hazard ratio function is
\begin{eqnarray}
\theta_h(t)
& = &  \theta^{id}_h(t) \left( \frac{ 1- \nu \log \left\{ S_0^{id}(t; 1) \right\}  }{1- \nu \log \left\{ S_1^{id}(t; 1) \right\}} \right).
\label{eq:popHR.as.idHR}
\end{eqnarray}
If a vaccine works such that $\theta^{id}_h(t)<1$
so that $S_1^{id}(t; 1) >   S_0^{id}(t; 1)$ for all $t>0$, then
$\theta_h(t) > \theta^{id}_h(t)$ (i.e., the population hazard ratio provides an upper bound on the
individual hazard ratio).

\begin{figure}[bt]
\includegraphics[width=6in]{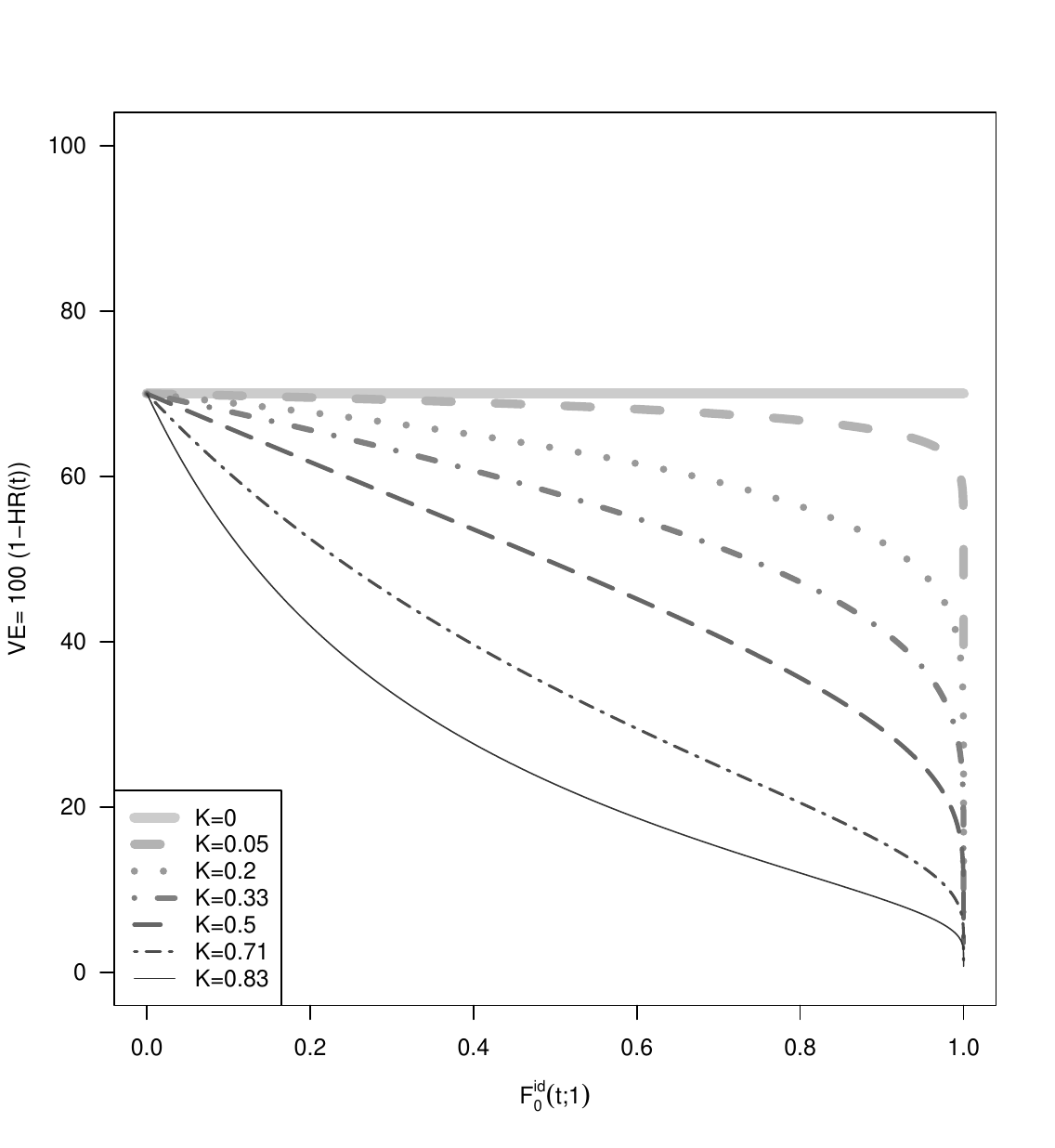}
\caption{Vaccine Efficacy by hazard ratio, $VE_h(t)$, under gamma frailty. Individual VE is 70\%,
lines are population VE estimands, $VE_{h}(t)$,  under different value for the variance of the gamma frailty distribution. We transform the variance of the frailty distribution, $var(U)=\nu$, to Kendall's tau, $K= \nu/(\nu+2)$. $K=0=\nu$ gives back the individual VE  of $VE^{id}_h=0.70$. The $x$-axis is the reference cumulative distribution for the control arm, $F_0^{id}(t;1)$.
\label{fig:VEwGammaFrailty}
 }
\end{figure}

If we can assume no vaccine waning so that $\phi_1(t)=\phi_1$ such that $\theta_{h}^{id}(t)=\theta_{h}^{id}=\phi_1/\phi_0$, then we have proportional individual-level hazards and the population hazard ratio from  equation~\ref{eq:popHR.as.idHR}
becomes,
\begin{eqnarray}
\theta_h(t)
& = &  \theta^{id}_h \left( \frac{ 1- \nu \log \left\{ S_0^{id}(t; 1) \right\}  }{1- \theta^{id}_h \nu \log \left\{ S_0^{id}(t; 1) \right\}} \right).
\label{eq:popHR.as.idHR.propHazards}
\end{eqnarray}
In Figure~\ref{fig:VEwGammaFrailty} we plot $VE_h(t)=1-\theta_h(t)$ under the no vaccine waning assumption when $\theta^{id}_h= 0.30$ with different values of $var(U_i)=\nu$.
To give the $var(U_i)$ interpretation, we can interpret them as Kendall's tau values (a type of rank correlation) between the $T_i(0)$ and $T_i(1)$ when there is no censoring (see  e.g., \citet{axelrod2023sensitivity} or
\citet{oakes1989bivariate}).  For the gamma frailty, Kendall's tau is $K = \frac{ \nu }{ \nu+ 2}$.

We see that at the population-level  $VE_h(t)$ is waning when  there is frailty (i.e., when
$var(U_i)>0$), even though we have $\theta_h^{id}(t)=\theta_h^{id}$ being constant over time.
The amount of population-level $VE_h(t)$ waning is smaller for lower values of $F_0(t)$ and lower values of $var(U)$ (and $K$).


\begin{figure}[bt]
\includegraphics[width=6in]{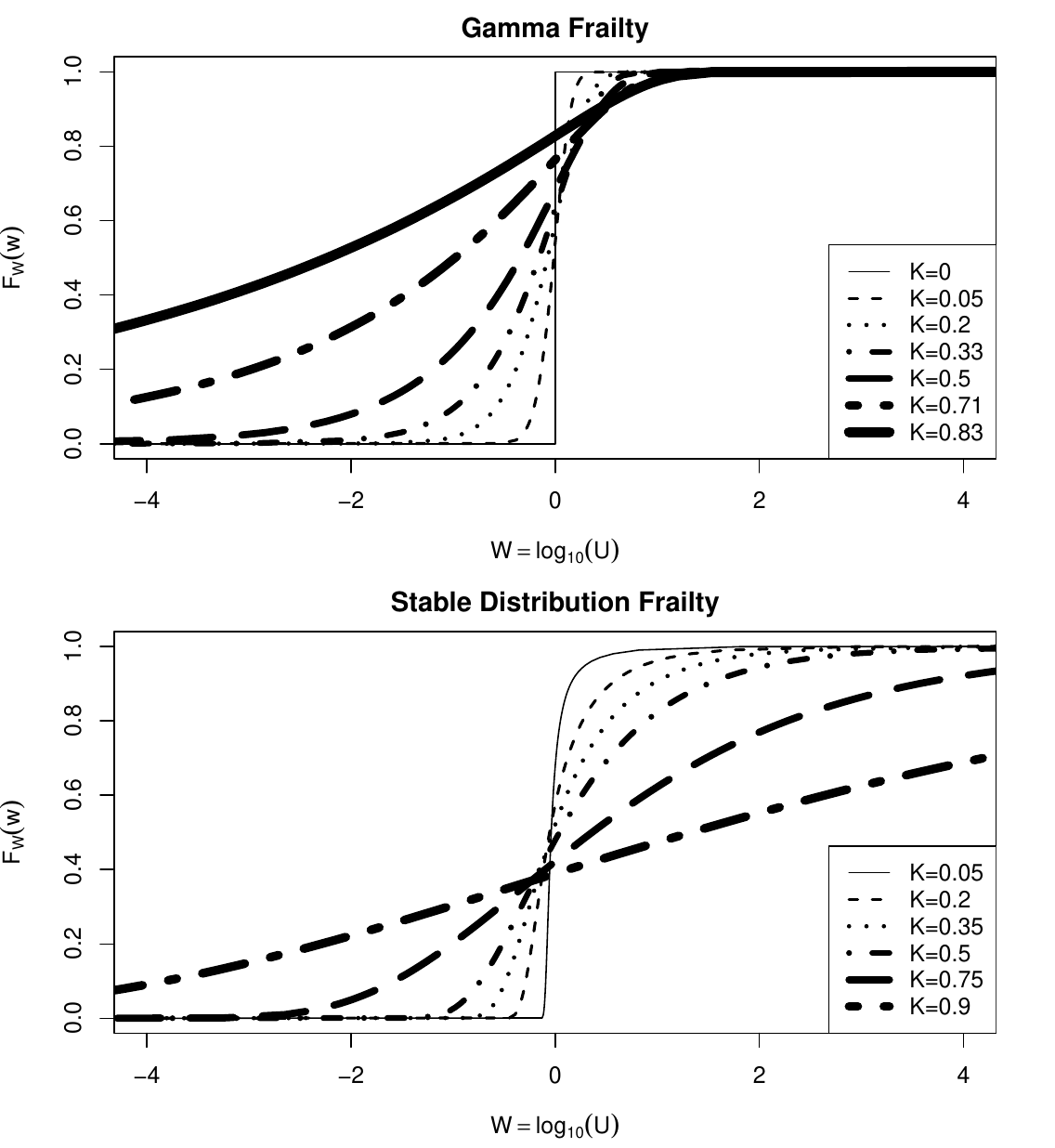}
\caption{ Plot of cumulative distributions of $W=log_{10}(U)$,
where $G$ is a frailty  random variable. The top panel is the gamma family of distributions,
all with mean 1 and different variances. The bottom panel is positive stable distributions
with parameters $PS(\alpha,\alpha,0)$
\citep[see][for stable distribution parameterization]{swihart2021bridged}.
\label{fig:PlotFrailtyDistributions}
 }
\end{figure}

In the top panel of Figure~\ref{fig:PlotFrailtyDistributions}
we plot the cumulative distributions of the $\log_{10}$ transformed frailty random variables,
 $log_{10}(U)$ for the gamma distributions with mean 1 and different variances (and hence different Kendall's tau values).
Because the mean is $1$, when the variance is close to zero, that has most of the density
clustered around $U=1$, which is $\log_{10}(U)=0$.  As the variance increases, because the mean remains
equal to $1$, most of the distribution occurs at $U<1$, which is $\log_{10}(U) < 0$.

Of course we can use other families for the frailty distribution. One interesting one is the
positive stable family of distributions \citep[see e.g.,][]{aalen2008survival,swihart2021bridged},
which when applied to individual-level proportional hazards models, returns population-level proportional hazards models but with different hazard ratios. The positive stable distribution that bridges between two proportional hazards can be
described with one parameter $0<\alpha<1$, which we write
$PS(\alpha,\alpha,0)$ \citep[see][for the positive stable distribution definition and parameterization]{swihart2021bridged}.
As with the gamma frailty, the positive stable frailty induces a Kendall's tau value, but here the relationship is $K=1-\alpha$ \citep{oakes1989bivariate}.
If the individual-level hazard ratio is $\theta_h^{id}$, and does not depend on time,
then the population hazard ratio is  $\theta_h = \left( \theta_h^{id} \right)^{\alpha}$.
(In the next section, we consider a parametric model with piecewise changing values of $\theta_h^{id}$ over time.)
For example, if the individual-level hazard ratio is $\theta^{id}=0.30$ as in Figure~\ref{fig:VEwGammaFrailty} giving $VE^{id}_h=70\%$, then the resulting population
vaccine efficacy for different values of $\alpha$ and $K$ would be,
$VE_h=70\% (\alpha \rightarrow 1, K=0),$
$68.1\% (\alpha=0.95, K=0.05),$
$61.8\% (\alpha=0.80, K=0.20),$
$54.3\% (\alpha=0.65, K=0.35),$
$45.2\% (\alpha=0.50, K=0.50),$
$26.0\% (\alpha=0.25, K=0.75),$ and
$11.3\% (\alpha=0.10, K=0.90).$
Unfortunately, the positive stable distribution does not have  a finite mean or variance,
so it is easiest to compare with the gamma distribution by examining the
cumulative distributions.  We plot the cumulative distributions of the $log_{10}$ transformed values of those positive stable distribution random variables in the bottom panel of Figure~\ref{fig:PlotFrailtyDistributions}.  Notice how likely extremely large values of the frailty random variables can be for small values of $\alpha$. Generally, as with the gamma frailty, as the Kendall's tau increases, the difference between the population-level $VE_h(t)$ and the individual-level $VE_h^{id}(t)=1-\theta_h^{id}(t)$ increases.


The main message of this section is to be careful interpreting population hazard ratios.
Suppose that the population $VE_h(t)$  follows the VE line with $K=0.33$ in Figure~\ref{fig:VEwGammaFrailty}.
 Then if the gamma frailty model with var$(U)=1$ (and $K=1/3$) is the true data generating model, then the apparent waning of the vaccine efficacy over time in terms of the population-level
 $VE_h(t)$, 
is not due to the vaccine becoming less effective for individuals over time (since under that model $VE_h^{id}(t)=0.70$ for all $t$), but it is due to the depletion of susceptibles effect caused by the frailty.
For real data, we do not know the data generating model. Another model that exactly fits the
VE line with var$(U)=1$ in Figure~\ref{fig:VEwGammaFrailty} could be generated by a homogeneous population. Under this second homogeneous model
the population-level $VE_h(t)$  equals the individual-level VE, so the VE is really waning on each individual.
For an application, we cannot differentiate between two models and must
rely on outside information to postulate which model is closer to reality.
For example, we can use individual antibody level when that can be a correlate of protection
as outside information  \citep[see e.g.,][]{gilbert2022immune}.

\section{Parametric Frailty with Ramp-Up Modeling}
\label{sec-parametric.ramp.up}

To tie these ideas of the last two sections to a real example,
consider the cumulative distribution curves from the BNT162b2 mRNA Covid-19 vaccine trial.
\citep[Figure 2 in: ][]{thomas2021safety}. The trial was a placebo-controlled trial
of over 44,000 participants, with over 93\% of them completing both doses of their vaccination
and entering the open-label follow-up.
 Participants were from several countries, and trial participants represent
a small portion of those country's populations, so as discussed in Section~\ref{sec-defining}, we expect there to be little interference.
 The vaccine efficacy was against laboratory-confirmed Covid-19 with an onset of 7 days or more after the second dose, which was given 21 days after the first dose.
 The vaccine efficacy estimand reported in \citet{thomas2021safety}
 was $VE_{IR}$ through 6 months of follow-up and was estimated
 as 91.3\% (95\% CI: 89.0, 93.2).

\begin{figure}[bt]
\includegraphics[width=6in]{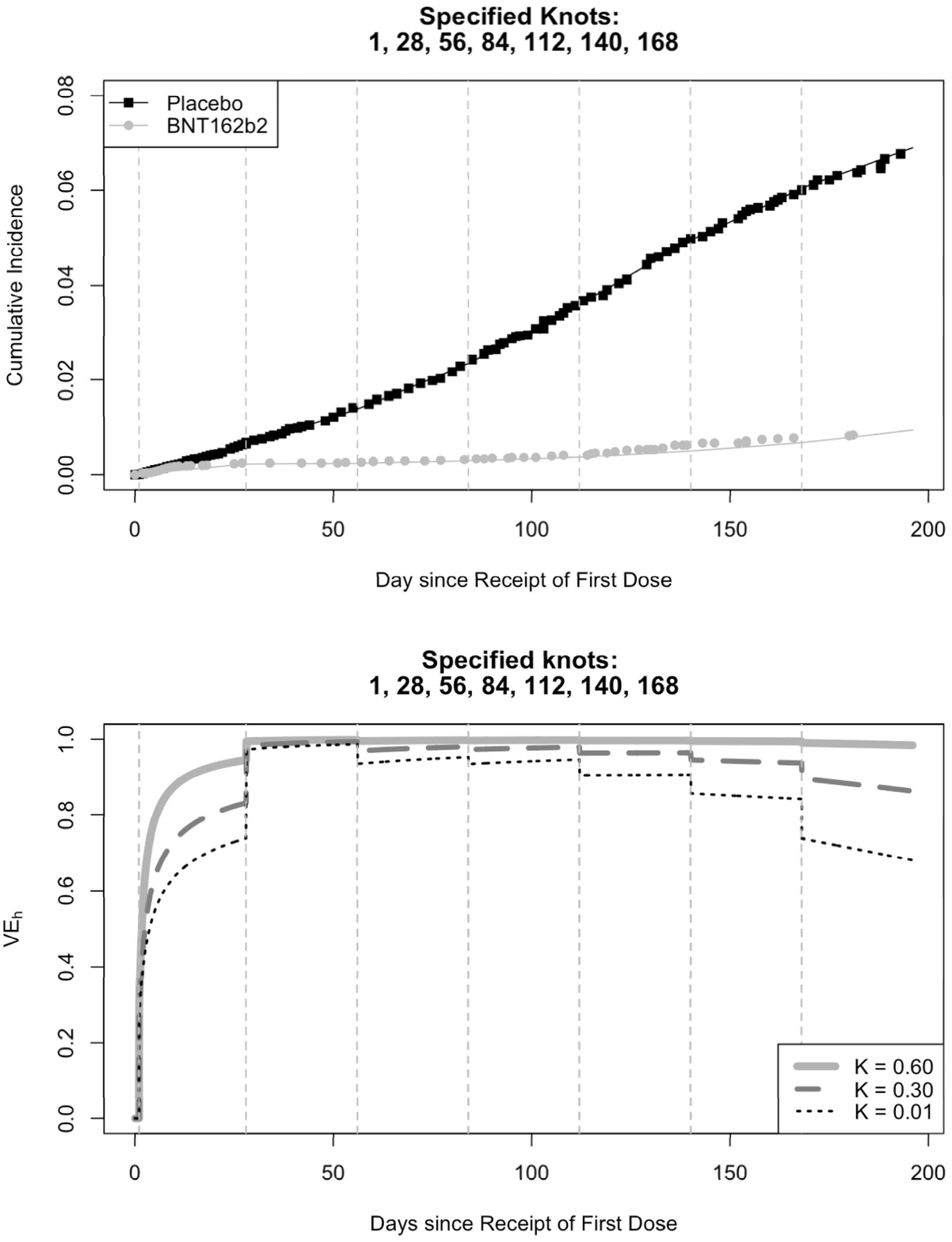}
\caption{ Piecewise Weibull model with positive stable frailty, fit to
an approximating dataset of the 
BNT162b2 mRNA Covid-19 vaccine trial \citep[Figure 2 in:][]{thomas2021safety}. Top panel is
estimated $F_0(t)$ and $F_1(t)$ values (points), with the model (lines).
Bottom panel is $VE_h^{id}(t)$ for different values of the positive stable distribution
parameter $\alpha=1-K$.
\label{fig:piecewiseWeibullwFrailty}
 }
\end{figure}

We use this example to describe the $VE_h(t)$ estimands to address some of the issues
of the last two sections. We had no access to the raw data
but approximate it by creating an individual patient data set using the {\sf IPDfromKM} R package \citep{liu2020IPDfromKM} and 
 Figure~2 of \citet{thomas2021safety}.
The {\sf IPDfromKM} package uses a point and click feature to manually trace digital Kaplan-Meier curves, and create a dataset to mimic those curves. 
The Kaplan-Meier estimates of $F_0(t)$ and $F_1(t)$  from the approximating dataset are given 
in the top panel of Figure~\ref{fig:piecewiseWeibullwFrailty}.
 The prespecified ramp-up period is about 28 days after the first dose, and we can see the day 14 values are close between the
two arms, suggesting at least a partial ramp-up during that period.
We fit a piecewise Weibull model with positive stable frailty
distribution, with knots at day 1 and then every  28 days until day 168. This is a flexible parametric model, where we force the vaccine efficacy to be $0$ at the first day, but in the other intervals we fit
Weibull models for each arm. The details of the model
 (including the R code and approximating dataset) 
are in
\citet{swihart2026fay}.
In the bottom panel
 of Figure~\ref{fig:piecewiseWeibullwFrailty}, 
we plot $VE_h^{id}(t)$
assuming different values of the frailty parameter $\alpha=1-K$. The value of $\alpha=0.99$ ($K=0.01$) is very close to
$VE_h(t)$. Because $VE_h(t)$ is measuring local hazards and allows jumping hazards, we can use it to
examine the changing population hazard ratio over time. The period from day 1 to day 28 represents most of the  pre-specified ramp-up period, and we see that the $VE_h(t)$ values are less that the later ones.
In later periods the $VE_h(t)$ values appear to slightly wane over time. For a sensitivity analysis,
we consider different values of $K$ (0.6 and 0.3), which show that if there is frailty then the $VE_h^{id}(t)$ are slightly higher.

\section{Discussion}
\label{sec-discussion}

We emphasize that for small event rates in the control arm, there is little
difference between estimands.
For example, consider a trial similar to
BNT162b2 mRNA Covid-19 vaccine trial \citep[Figure 2 in:][]{thomas2021safety}.
Suppose we know the true distributions at day 182, that $\tau=182$,  and
${F}_0(\tau) = 0.065$ and ${F}_1(\tau) = 0.008$.
Then the corresponding ITT VE estimands are
$VE_{CI}(\tau)= 87.7\% <$ $VE_{CH}(\tau)= 88.0\% <$ $VE_{odds}(\tau)= 88.4\%$, 
and by equation~\ref{eq:thetaIR.inequalities} we have
$87.6\% \leq VE_{IR}(\tau) \leq 88.5\%$.
The value of $VE_{Cox}(\tau)$ cannot be calculated using only $F_0(\tau)$ and $F_1(\tau)$, but if the distributions have approximately proportional hazards,
then  $VE_{Cox}(\tau) \approx$
$VE_{CH}(\tau)=88.0\%$.

An important point of this paper is that the ITT VE estimands may be prespecified and defined
nonparametrically
(e.g., do not depend on assumptions like proportional hazards). Then for small control event rates, since the VE estimands are similar, we are free to choose the ones with simple straightforward interpretation like $VE_{IR}(\tau)$ or $VE_{CI}(\tau)$.
The estimand $VE_{Cox}(\tau)$ is difficult to interpret precisely
when the proportional hazards assumption fails
because in this context, our verbal interpretation of the estimand must explain the complex weighted average of hazard ratios and how the weighting scheme depends on the value of the estimand itself.
Under proportional hazards $VE_{Cox}(\tau)=VE_{CH}(\tau)$.
The estimand $VE_{CH}(\tau)$ provides an alternative to $VE_{Cox}(\tau)$ that is equal to the Cox estimand under proportional hazards, but has a simpler interpretation when proportional hazards fails.
 When proportional hazards approximately holds (and hence $VE_{Cox}(\tau) \approx VE_{CH}(\tau)$), then the Cox model may provide an efficient estimator of $VE_{CH}(\tau)$.

The choice between the ITT VE estimates becomes more important as $F_0(\tau)$ becomes larger (see Figure~\ref{fig:VE2plots}). Then researchers should be aware that $VE_{CI}(\tau) < VE_{CH}(\tau) < VE_{odds}(\tau)$, and all the ITT VE estimands are not interchangeable. Further, there are simple
formulas for converting some VE estimands to other ones (see equations~\ref{eq:theta.CI.into.theta.CH} and \ref{eq:thetaCI.using.thetaOdds}).

By tradition, the full immunization VE estimands are preferred for the primary endpoint, but we have shown that those estimands may give misleading inferences in extreme cases where the vaccine actually increases
the event rate compared to control. Although  extreme cases like  Figure~\ref{fig:RampUpCumHaz} Scenario~3
are unlikely, by defining an estimand that can show a positive effect for a vaccine that hurts a population (in the sense that $F_1(\tau)>F_0(\tau)$), there is room for smaller errors of interpretation where a full immunization VE estimand looks slightly better than it should because of early vaccine negative effects. Since traditions are hard to change, even when using the full immunization VE estimand as the primary endpoint estimand,  it is important to plot the distributions
and useful to estimate an ITT VE estimand as well.

Finally, we have shown that when examining the durability of a vaccine from  randomized trial results only, the population $VE_h(t)$ curves should not be interpreted as individual-level
local vaccine efficacy effects. Figure~\ref{fig:VEwGammaFrailty} shows that if the within individual correlation of the potential outcomes is large, or $F_0(\tau)$ is large, then frailty can induce
an apparent waning in the population $VE_h(t)$ curve. A piecewise Weibull frailty model can explore some of those effects (see Figure~\ref{fig:piecewiseWeibullwFrailty}).

\begin{appendix}

\section{Weighted Average Discrete Hazard}
\label{sec-weighted.avg.discrete.haz}

The methods in this paper deal with continuous time.
Here we describe a discrete time method and show its relationship to some continuous time models.

We can define an average of discrete hazards. To relate to the continuous models, we partition the continuous space into disjoint sets.
Suppose we assess participants at regular times, $0=t_0 <t_1<t_2 \ldots$,
and define the $j$th time period as $(t_{j-1},t_j]$.  Let
$p_0(t_j)= F_0(t_j) - F_0(t_{j-1})$ be the proportion of the control population who get an event during the $j$th time period.
Let the discrete hazard $h_0(t_j)$ be the proportion of the control population {\it who are at risk during the
$j$th time period}  that get the event during that period,
\begin{eqnarray*}
h_0(t_j) & \equiv &  Pr[ T_i(0) \in (t_{j-1},t_j] | T_i(0) > t_{j-1} ] \\
       & = &  \frac{ p_0(t_j) }{ 1 - F_0(t_{j-1}) } = \frac{p_0(t_j) }{ S_0(t_{j-1}) }.
\end{eqnarray*}
Similarly define $F_1(t_j)$, $S_1(t_j)$, $p_1(t_j)$, and $h_1(t_j)$ for the vaccine arm.
Since for continuous data, $S_z(t) = \exp\ \left\{- \int_0^{t} \lambda_z(u) du \right\}$, we can rewrite
$h_z(t_j)$ as
\begin{eqnarray*}
h_z(t_j) & = & \frac{ S_z(t_{j-1}) - S_z(t_j) }{ S_z(t_{j-1})} = 1 - \exp \left( - \int_{t_{j-1}}^{t_j} \lambda_z(t) dt \right).
\end{eqnarray*}

Suppose the assessment times are equally
spaced
such that $t_j-t_{j-1}=\Delta$ for all $j$.
For smooth hazards and small values of  $\Delta$, we have $\int_{t_{j-1}}^{t_j} \lambda_z(t) dt \approx \Delta \lambda(t_j)$. Further, we can show  by Taylor series methods that
 for small $x$, then $x \approx 1- \exp(-x)$. Combining those approximations, for small $\Delta$ and smooth hazards, we have
\begin{eqnarray*}
  h_z(t_j) & \approx & 1 - \exp( - \Delta \lambda_z(t_j) ) \approx \Delta \lambda_z(t_j),
\end{eqnarray*}
and
\begin{eqnarray*}
\theta_{dh}(t_j) & = & \frac{ h_1(t_j) }{ h_0(t_j) }  \approx \frac{ \lambda_1(t_j) }{ \lambda_0(t_j) } = \theta_h
\end{eqnarray*}
A weighted average of the discrete hazards up to time $\tau = t_k$ is
\begin{eqnarray}
\theta_{wdh}(\tau; k)  & = & \frac{ \sum_{j=1}^{k} w_j \theta_{dh}(t_j) }{ \sum_{j=1}^{k} w_j  }.
\label{eq:theta.wdh.tau.k}
\end{eqnarray}

Now suppose we fix $\tau$, use equally spaced assessments, but allow the number of assessments
to go to $\infty$.
Suppose there are $k$ assessments after baseline up to $\tau$, $t_0 < t_{1,k} < \cdots < t_{k,k}=\tau$.
Let
\begin{eqnarray*}
w(t;k) = \sum_{j=1}^{k} w_{j,k} I(t_{j-1,k} < t \leq t_{j,k}),
\end{eqnarray*}
such that $\lim_{k \rightarrow \infty} w(t;k) \equiv w(t)$.
Then
\begin{eqnarray}
\lim_{k \rightarrow \infty} \theta_{wdh}(\tau; k)  & = & \lim_{k \rightarrow \infty} \frac{ \sum_{j=1}^{k} \int_{t_{j-1}}^{t_j} w_{j;k} \theta_{dh}(t_j) dt  }{ \sum_{j=1}^{k} \int_{t_{j-1}}^{t_j} w_{j,k} dt } \nonumber \\
&  = &  \frac{ \int_{0}^{\tau} w(t) \theta_h(t) dt }{ \int_{0}^{\tau} w(t)  dt } = \theta_{wh}(\tau),
\label{eq:lim.theta.wdh}
\end{eqnarray}
When $t_1=\tau$ (i.e., we only assess at the end of the study), then $h_z(t_1) = F_z(t_1)$
and $\theta_{wdh}(\tau; k)  = \theta_{CI}(\tau)$.

To compare the average discrete hazard ratio to the continuous hazard ratio, consider the proportional hazards model
with the distribution of the control arm exponential. Then for the continuous hazard ratio,
 $\theta_h=\theta_{CH}(\tau)=\lambda_1/\lambda_0$, where $\lambda_z$ is the rate parameter
in the exponential distributions of the $z^{th}$ arm.
Further, if we have equally spaced assessments on the discrete hazard ratio, then
$h_z(t_i)=h_z(t_j)= h_{z; k}$ for all $i,j$.
We can derive an expression for $h_{z;k}$ in terms of $\lambda_z$:
\begin{eqnarray*}
& & S_z(\tau) = \exp( - \lambda_z \tau) = \prod_{j=1}^{k} (1 - h_{z;k}) =(1 - h_{z;k})^k  \\
\Rightarrow & & 1- h_{z;k} =\left\{ \exp( - \lambda_z \tau) \right\}^{1/k}   \\
\Rightarrow & & h_{z;k} = 1 - \exp \left(  - \frac{\lambda_z \tau}{k} \right)  \\
\end{eqnarray*}

In Table~\ref{tab-VECH.vs.VEdh} we show the difference between $VE_{wdh}(\tau;k)$
and $VE_{CH}(\tau)$ in the two exponential case. This case is a proportional hazards case
so $VE_{Cox}(\tau) = VE_{CH}(\tau)= VE_{wh}(\tau)$ for any weights, and  with equally spaced
assessments  $VE_{wdh}(\tau;k)$ does not depend on weights either.
The biggest difference is when $k=1$,
which is the same as $VE_{CI}(\tau)$, so we will not explore $VE_{dh}(\tau; k)$ further.

\begin{table}
\caption{Comparison of $VE_{CH}(\tau)$ to $VE_{dh}(\tau; k)$ under different values of $k$ and $F_0(\tau)$.
Suppose we have a 52 week study (about 1 year), and we consider 4 different assessment schedules, yearly (once at the end of the study), every 4 weeks, weekly, and daily.  When $VE_{CH}(\tau)= 50\%$ we give   $VE_{dh}(\tau; k)$  for two different values of $F_0(\tau)$
\label{tab-VECH.vs.VEdh}
}

\begin{tabular}{lrccc}
\hline
  &   &  $VE_{CH}(\tau)$  &    $VE_{dh}(\tau; k)$   & $VE_{dh}(\tau; k)$ \\
Schedule & k & &  ($F_0(\tau)=10\%$) & ($F_0(\tau)=20\%$) \\ \hline
yearly & 1 &   $50.0\%$ &     $48.7\%$ & $47.2\%$ \\
quarterly & 4 &   $50.0\%$ &   $49.7\%$ & $49.3\%$ \\
every 4 wks & 13 &   $50.0\%$ &   $49.9\%$ & $49.8\%$ \\
every week & 52 &   $50.0\%$ &   $50.0\%$ & $49.9\%$ \\
daily & 364 &   $50.0\%$ &   $50.0\%$ & $50.0\%$ \\ \hline
\end{tabular}


\end{table}

\section{Peak VE Estimand Differences Equivalence}
\label{app-peakVEdiff.equivalence}

We want to show that $\Delta_{1max}(F_0(\tau))=\Delta_{2max}(F_0(\tau))$, where
 $\Delta_{1max}(F_0(\tau))$ is
 the maximum value of
\begin{eqnarray*}
\Delta_1 \left\{ F_0(\tau),F_1(\tau) \right\} & = & VE_{CH}(\tau) - VE_{CI}(\tau) \\
& = & \frac{F_1(\tau)}{F_0(\tau)} - \frac{ \log(1-F_1) }{\log(1-F_0(\tau)}
\end{eqnarray*}
for fixed $F_0(\tau)$ over different values of $F_1(\tau)$ such that $VE_{CI}(\tau) \in (0,1)$,
and
 $\Delta_{2max}(F_0(\tau))$ is
 the maximum value of
\begin{eqnarray*}
\Delta_2 \left\{ F_0(\tau),F_1(\tau) \right\} & = & VE_{odds}(\tau) - VE_{CH}(\tau) \\
& = &  \frac{ \log(1-F_1) }{\log(1-F_0(\tau)} - \frac{F_1(\tau) \left\{ 1- F_0(\tau) \right\} }{ F_0(\tau) \left\{ 1- F_1(\tau) \right\} }
\end{eqnarray*}
for fixed $F_0(\tau)$ over different values of $F_1(\tau)$ such that $VE_{CH}(\tau) \in (0,1)$.

For notational ease, in this section we write $F_0 \equiv F_0(\tau)$ and $F_1 \equiv F_1(\tau)$.
Further, we note that for fixed $F_0$, $F_1 \in (0, F_0)$ is equivalent to $VE_{CI}(\tau) \in (0,1)$
or $VE_{CH}(\tau) \in (0,1)$.

To find $\Delta_{1max}(F_0)$ we calculate
\begin{eqnarray*}
\frac{ \partial \Delta_1(F_0,F_1) }{\partial F_1} & = &  \frac{1}{F_0} + \frac{ 1}{(1-F_1) \log(1-F_0) }
\end{eqnarray*}
set it to $0$, and solve for $F_1$ to get
\begin{eqnarray*}
F_{1max} & = & 1 + \frac{ F_0 }{ \log(1-F_0) }
\end{eqnarray*}
so that
\begin{eqnarray*}
\Delta_{1max}(F_0) &  = & \Delta_1(F_0,F_{1max}) \\
& = &  \frac{F_{1max}}{F_0} - \frac{ \log(1-F_{1max}) }{\log(1-F_0) } \\
& = & \frac{1}{F_0} + \frac{1}{\log(1-F_0)}  -  \frac{
\log \left\{ \frac{-F_0}{\log(1-F_0)} \right\} }{\log(1-F_0)}
\end{eqnarray*}

To find $\Delta_{2max}(F_0)$ we calculate
\begin{eqnarray*}
\frac{ \partial \Delta_2(F_0,F_1) }{\partial F_1}
& = &  \frac{ -1}{(1-F_1) \log(1-F_0) }-  \\
& & \frac{(1-F_0)}{F_0} \left( \frac{1}{(1-F_1)^2 }    \right)
\end{eqnarray*}
set it to $0$, and solve for $F_1$ to get
\begin{eqnarray*}
F_{1max}^* & = & 1 - \log(1-F_0) + \frac{\log( 1-F_0) }{F_0}  \\
& = & 1 + \log(1-F_0) \left( -1 + \frac{1}{F_0} \right) \\
& = & 1 + \log(1-F_0) \left( \frac{1-F_0}{F_0} \right)
\end{eqnarray*}
so that
\begin{eqnarray*}
\frac{F_{1max}^*}{1-F_{1max}^*} & = & \frac{1 + F_0^{-1} (1-F_0) \log(1-F_0) }{-F_0^{-1} (1-F_0) \log(1-F_0)} \\
& = & \frac{-F_0 - (1-F_0) \log(1-F_0) }{(1-F_0) \log(1-F_0)} \\
& = & -1 - \frac{F_0}{(1-F_0)\log(1-F_0) }
 \end{eqnarray*}
and
\begin{eqnarray*}
\Delta_{2max}(F_0) &  = & \Delta_2(F_0,F_{1max}^*) \\
& = &   \frac{ \log(1-F_{1max}^*) }{\log(1-F_0) }  - \frac{ (1-F_0) F_{1max}^* }{ F_0 (1-F_{1max}^*) } \\
& = &  \frac{ \log \left( - \log(1-F_0) \left( \frac{1-F_0}{F_0} \right) \right) }{\log(1-F_0) }  - \\
& & \left\{ - \frac{ (1-F_0) }{ F_0 }  -  \frac{1 }{ \log(1-F_0) } \right\}  \\
& = &  \frac{ \log(1-F_0) +  \log \left( - \frac{ \log(1-F_0) }{F_0} \right) }{\log(1-F_0) }  + \\
& & \left\{ -1 + \frac{1 }{ F_0 }  + \frac{1 }{ \log(1-F_0) }  \right\} \\
& = &  1 + \frac{ - \log \left( \frac{ -F_0}{\log(1-F_0) } \right) }{\log(1-F_0) }  + \\
& &  -1 + \frac{1 }{ F_0 }  +  \frac{1 }{ \log(1-F_0) }   \\
& = &   \frac{ - \log \left( \frac{ -F_0}{\log(1-F_0) } \right) }{\log(1-F_0) }  + \\
& &  \frac{1 }{ F_0 }  +  \frac{1 }{ \log(1-F_0) },
\end{eqnarray*}
which is the same as
$\Delta_{1max}(F_0).$

\section{Details of Figure~\ref{fig:RampUpCumHaz} }
\label{app:RampUpCumHaz}

 In all 3 scenarios, the end of the ramp-up time is $t_{RU}=28$, the end of the study is
$\tau = 150$, and the time for the control vaccine is distributed exponential with
rate $\lambda_0=0.0005$, so that $F_0(t_{RU}) \approx 0.014$ and $F_0(\tau) \approx 0.072$.
Here are the details of the test vaccine distributions.
\begin{description}
\item[Scenario~1:] A ramp-up model with  $VE_{CH}^*(t^*)=0.70$ for all $t^*>0$. \\
Specifically,
\begin{eqnarray*}
\lambda_1(t) = \left\{
\begin{array}{cc}
\lambda_0 &  t \leq t_{RU}  \\
0.30 \lambda_0 & t > t_{RU}
\end{array}
\right.
\end{eqnarray*}
so that
$F_1(t)=F_0(t)$ for $t \leq t_{RU}$, and for $t> t_{RU}$,
\begin{eqnarray*}
F_1(t) &= & F_1^*(t-t_{RU}) \left\{ 1 - F_1(t_{RU}) \right\} + F_1(t_{RU}),
\end{eqnarray*}
where $F_1^*(t-t_{RU})= F_1^*(t^*)$ is exponential with rate $0.30 \lambda_0$.
\item[Scenario~2:] A model where $\theta_{h}(t)=\lambda_1(t)/\lambda_0$ goes from $\theta_h(0)=\psi_1$ to
$\theta_h(t_{RU})=\psi_2$, and then remains $\theta_h(t)=\psi_2$ for all $t > t_{RU}$. \\
Specifically, for $t \leq t_{RU}$ the hazard for the test vaccine is
\begin{eqnarray*}
\lambda_1(t) & = & \lambda_0 \left\{ \psi_1  - (\psi_1 -\psi_2)  \left( \frac{t}{t_{RU}} \right) \right\},
\end{eqnarray*}
 and for $t > t_{RU}$ the hazard is
$\lambda_1(t)=\lambda_0 \psi_2$. Then
\begin{eqnarray*}
F_1(t) & = &  1 - \exp \left( - \int_0^t \lambda_1(u) du \right) \\
& = & \left\{
\begin{array}{l}
1 - \exp \left( - \lambda_0 \psi_1 t +   \frac{t^2 (\psi_2-\psi_1) \lambda_0 }{2 t_{RU}} \right) \\
\hspace*{10em} \mbox{ if } t \leq t_{RU} \\
1 - (1-F_1(t_{RU})) \exp \left(  - \psi_2 \lambda_0 (t - t_{RU}) \right) \\
\hspace*{10em} \mbox{ if } t > t_{RU}
\end{array}
\right.
\end{eqnarray*}
Since $1-F_1^*(t^*) = \frac{1-F_1(t)}{1-F_1(t_{RU})}$, then
$1-F_1(t^*) = \exp \left(  - \psi_2 \lambda_0 t^* \right)$, which is exponential with hazard
$\psi_2 \lambda_0$. Further, because of the memoryless property of the exponential, $1-F_0(t^*)= \exp \left(  - \lambda_0 t^* \right)$ is exponential with hazard $\lambda_0$. So $VE_{CH}^{*}(t^*) = 1-\psi_2$.
Scenario 2 has $\psi_1=1$ and $\psi_2=0.30$.
\item[Scenario~3:] Same as Scenario~2, except with  $\psi_1=3$ and $\psi_2=0.70$.
\end{description}

\end{appendix}

\bibliographystyle{unsrtnat}
\bibliography{refs}

\end{document}